\documentclass[%
aip,
jcp,
amsmath,amssymb,
]{revtex4-2}

\usepackage[table]{xcolor}
\usepackage{nicematrix}
\usepackage[utf8]{inputenc}
\usepackage{amsmath, amssymb}
\usepackage{physics}
\usepackage{tikz}
\usetikzlibrary{arrows.meta}
\usepackage{subcaption}
\usetikzlibrary{decorations.pathreplacing}
\usepackage[export]{adjustbox}
\usepackage{makecell}
\usepackage{pgf}
\usepackage{array}

\usepackage{tabularx}
\usepackage{xcolor}
\usepackage{multirow}
\usepackage{booktabs}
\newcommand{\irr}[2]{\mathrm{#1}_{\mathrm{#2}}}

\usepackage[utf8]{inputenc}
\usepackage[T1]{fontenc}
\usepackage{graphicx}
\usepackage{dcolumn}
\usepackage{bm}
\usepackage{float}
\usepackage{lipsum}
\usepackage{siunitx}
\usepackage[colorlinks=true, citecolor=blue, linkcolor=black, urlcolor=blue, pdfborder={0 0 0}]{hyperref}
\usepackage{caption}
\usepackage{subcaption}

\usepackage[capitalise]{cleveref}

\tikzset{
    pics/full/.style={
        code={
            \draw (0,0.5) -- (1.25,0.5);
            \draw[-{Stealth[left]}, very thick] (0.5,0) -- (0.5,1.0);
            \draw[-{Stealth[left]}, very thick] (0.75,1.0) -- (0.75,0);
        }
    },
    pics/up/.style={
        code={
            \draw (0,0.5) -- (1.25,0.5);
            \draw[-{Stealth[left]}, very thick] (0.5,0) -- (0.5,1.0);
        }
    },
    pics/down/.style={
        code={
            \draw (0,0.5) -- (1.25,0.5);
            \draw[-{Stealth[left]}, very thick] (0.75,1.0) -- (0.75,0.0);
        }
    }
}

\newlength{\myfigwidth}

\newcommand{\buplus}[1]{$#1^1\mathrm{B}_{\mathrm{u}}^{+}$}

\newcommand{\agminus}[1]{$#1^1\mathrm{A}_{\mathrm{g}}^{-}$}
\newcommand{\buminus}[1]{$#1^1\mathrm{B}_{\mathrm{u}}^{-}$}

\usepackage{tikz}

\begin{document}

\title{
Geometry-Dependent Electronic Coupling and Singlet Fission Pathways in Carotenoid Dimers}

\author{Alexandru G. Ichert}
\email{alexandru.ichert@chem.ox.ac.uk}
\affiliation{Department of Chemistry, Physical and Theoretical Chemistry Laboratory, University of Oxford, Oxford, OX1 3QZ, United Kingdom}
\affiliation{Linacre College, University of Oxford, Oxford, OX1 3JA, United Kingdom}

\author{George A. Smith}
\affiliation{Department of Chemistry, Physical and Theoretical Chemistry Laboratory, University of Oxford, Oxford, OX1 3QZ, United Kingdom}
\affiliation{Worcester College, University of Oxford, Oxford, OX1 2BH, United Kingdom}

\author{William Barford}
\email{william.barford@chem.ox.ac.uk}
\affiliation{Department of Chemistry, Physical and Theoretical Chemistry Laboratory, University of Oxford, Oxford, OX1 3QZ, United Kingdom}

\date{\today}

\begin{abstract}

Using the density matrix renormalization group method, we solve the Pariser--Parr--Pople model of $\pi$-electron systems to determine the character and energies of the low-lying eigenstates of a lycopene dimer comprising two 22-carbon-atom chains. To complement these calculations, we use the molecular-exciton, valence-bond model of polyene systems to derive selection rules governing interchain coupling, which provide a physical framework for interpreting the DMRG results.
We find that dimer geometry strongly influences both the character and energetics of the low-lying eigenstates. In vertically stacked dimers, neither the bright-dimer state nor the triplet-pair states contain any interchain charge-transfer (CT) character. In contrast, in slip-stacked dimers, both the bright-dimer state and triplet-pair states do contain interchain CT character. We also find that increasing the interchain coupling preferentially stabilizes states with greater interchain character, leading to a reordering of the low-lying eigenstates.
These results have direct implications for energetically possible singlet fission (SF) mechanisms in carotenoid aggregates. For vertically stacked dimers, the only energetically viable pathway identified here proceeds through intermediate intrachain triplet-pair states, whereas slip-stacked geometries  allow an additional SF pathway mediated by interchain CT states. Our results establish a direct connection between carotenoid packing geometry, symmetry-controlled electronic coupling, and the energetically accessible pathways for SF.

\end{abstract}

\maketitle

\section{Introduction}\label{ch:1}

Singlet fission (SF) is a photophysical process in which a photoexcited singlet state generates a pair of spin-correlated triplet excitations with an overall singlet character, which may then decohere into independent triplets.\cite{Smith2010,Smith2013} Observed first in anthracene crystals\cite{Singh1965} and subsequently in tetracene,\cite{Swenberg1968} SF has attracted considerable interest because of its potential to overcome the  theoretical efficiency limit of single-junction photovoltaic devices.\cite{Shockley1961,Hannah2006} Extensive experimental and theoretical work has aimed to determine the microscopic mechanism of SF in acenes, with  substantial evidence supporting an important role for intermolecular charge-transfer (CT) states.\cite{Berkelbach2013,Aryanpour2015,Monahan2015} In contrast, the mechanism of SF in carotenoids remains considerably less well understood.

The electronic states of carotenoids and the closely related polyenes have been the subject of intense study for over five decades.\cite{Schulten1972,Schulten1976,Hayden1986,Chandross1999,Barford2022,WBBook,Valentine2020} Their accurate description remains challenging because of the strong multireference character of their low-lying eigenstates. An important unresolved issue is the relative ordering of the vertical excitation energies of the bright $1^1\textrm{B}_\textrm{u}^+$ and dark $2^1\textrm{A}_\textrm{g}^-$ states. 
Taffet \textit{et al.}\cite{Taffet2019} predict that the vertical  \buplus{1} energy lies below the vertical  \agminus{2} energy for  carotenoids with fewer than 26 carbon atoms, with a near degeneracy at 26 carbon atoms. Similarly, Khokhlov and Belov\cite{khokhlov_ab_2020} predict that the vertical  \buplus{1} energy lies below the vertical  \agminus{2} energy for polyenes with  between 16 and 26 carbon atoms, with a reversal of their relaxed energies. 
Multireference configuration-interaction singles-doubles with perturbative corrections on polyenes by Chagas \textit{et al.}\cite{Chagas2025} predict that the vertical \buplus{1} energy lies below the vertical \agminus{2} energy by ca.\ 0.1 eV for polyenes of up to 10 carbon atoms, with a reversal of the ordering for longer polyenes. The ordering of the dark $2^1\textrm{A}_\textrm{g}^-$ state  below the bright $1^1\textrm{B}_\textrm{u}^+$ state at the Franck-Condon point is commonly assumed for polyenes.\cite{Aryanpour2015,Prodhan2017,Valentine2020,Santra2022}

Direct experimental determination of this ordering is complicated both by the predominantly doubly excited character of the $2^1\textrm{A}_\textrm{g}^-$ state, rendering it optically inaccessible from the ground state,\cite{Schulten1972,Starcke2006,Boguslawski2019,Chagas2025,Ichert2025} and by interchain coupling that blue-shifts the bright-dimer state relative to the intrachain dark triplet-pair states.

In carotenoids and polyenes in solution, internal conversion from the bright $1^1\textrm{B}_\textrm{u}^+$ state to the dark $2^1\textrm{A}_\textrm{g}^-$ state occurs on a subpicosecond timescale.\cite{Musser2015,Quaranta2021,Manawadu2022,Manawadu2023,Peng2024,Georges2026} SF, by contrast, has been observed in biological systems\cite{Portoles2018,Quaranta2021,Wang2026} and synthetic aggregates,\cite{Wang2010,Musser2015,Kundu2021,Peng2024,Magne2025} but not in isolated carotenoid molecules. SF in multichromophoric environments, together with its absence in isolated molecules, therefore  suggests that carotenoid SF is an intrinsically bimolecular process.

Despite considerable effort, no experimental consensus has emerged  for the microscopic pathway by which the photoexcited bright state evolves into a correlated singlet triplet-pair in carotenoids. Several experimental studies have proposed a mechanism mediated by an intrachain triplet-pair intermediate belonging to the $2\textrm{A}_\textrm{g}$ family of states,\cite{Portoles2018,Kundu2021,Valentine2020,Magne2025} whereas other evidence suggests the involvement of an intermolecular CT intermediate.\cite{Peng2024} A third possibility is direct formation of the correlated triplet-pair from the bright state, without an intermediate state.\cite{Musser2015} 

A theoretical description of carotenoid SF has also proved  challenging. The rapid growth of the many-electron Hilbert space with the number of coupled chains has restricted most studies to dimers\cite{Prodhan2017,Santra2022,Santra2025} or, at most, four short chains.\cite{Aryanpour2015}
Using up to four chains of 8 carbon atoms, Aryanpour \textit{et al.}\cite{Aryanpour2015} investigated the effects of aggregation on the energies and character of the resulting eigenstates. Their results suggest that CT states are a plausible intermediate for SF in carotenoid aggregates, but a mechanism involving intrachain triplet-pair states is also possible.
~Santra \textit{et al.}\cite{Santra2022,Santra2025} analyzed couplings between diabatic states of decapentaene dimers (i.e., $2\times10$ carbon atoms). They found that SF can, in principle, proceed through either CT or $2\textrm{A}_\textrm{g}^-$ intermediates, with the preferred pathway depending on the aggregate geometry. This geometry-dependent picture was used to rationalize the distinct SF behavior observed in different astaxanthin aggregates.\cite{Musser2015}
Finally, a mechanism mediated solely by intrachain triplet-pair states has also been proposed by Barford \textit{et al.},\cite{Chambers2023,Barford2023} and subsequently used to investigate spin decoherence of the entangled triplet-pair.\cite{marcus_triplet-triplet_2020, Ichert2025} 

Resolving these proposed competing mechanisms in carotenoid systems requires a detailed understanding of the character and energies of their low-lying eigenstates. Here, we address this problem using two complementary approaches. First, we obtain numerically accurate solutions of the Pariser-Parr-Pople (PPP) $\pi$-electron model for lycopene dimers (of two chains of  $22$ carbon atoms) using the density matrix renormalization group (DMRG) method. We characterize the  dimer eigenstates by projecting their matrix product state (MPS) representations onto products of the eigenstates of the individual chains. Second, we employ a molecular-exciton, valence-bond model of polyene systems to construct a coarse-grained description of the dimer excited states. Within this framework, we derive symmetry-based selection rules governing interchain coupling between the basis states, providing a  physical interpretation of our DMRG results.

Our key result is that dimer conformation controls the character and energetics of the low-lying excited states and, consequently, the energetically accessible pathways for SF. For vertically stacked dimers (illustrated in Fig.~1(a)), symmetry prevents the bright-dimer and low-lying triplet-pair states from acquiring interchain CT character, leaving an intrachain triplet-pair-mediated pathway as the only energetically viable mechanism. Slip-stacking (illustrated in Fig.~1(b)) breaks this symmetry and allows coupling to interchain CT states, thus opening an additional CT-mediated pathway. Increasing the interchain coupling further stabilizes states with greater interchain character and can reorder the low-lying eigenstates. These results establish a direct connection between molecular packing, symmetry-controlled electronic coupling, and the energetically accessible pathways for SF in carotenoid aggregates.


The remainder of this paper is organized as follows. In Sections~\ref{ch:2} and~\ref{ch:3}, we introduce the model and its parameters, and define the two dimer geometries considered: vertically stacked and slip-stacked chains. Section~\ref{ch:4} briefly describes the DMRG methodology. In Section~\ref{RM}, we develop a reduced model for the electronic states of polyene dimers and derive selection rules that govern  interchain coupling. We present our results in Section~\ref{ch:6} and summarize our conclusions in Section~\ref{ch:7}. The Supporting Information (SI) contains additional information on the DMRG method and the use of the molecular-exciton, valence-bond model to derive the symmetry-based selection rules.

\section{$\pi$-Electron Model}\label{ch:2}

The PPP Hamiltonian for a linear chain of $N$ conjugated carbon atoms is
\begin{widetext}
\begin{eqnarray}\label{eq:PPP_Ham}
    \hat{H}_{\textrm{PPP}}= && -\sum_{i \sigma}^{N-1}t_i \left( \hat{c}_{i\sigma}^\dagger \hat{c}_{i+1\sigma} + \hat{c}_{i+1\sigma}^\dagger \hat{c}_{i\sigma} \right)
    + U\sum_{i}^{N} \left( \hat{N}_{i\uparrow}-\frac{1}{2} \right)\left( \hat{N}_{i\downarrow}-\frac{1}{2} \right)\\ \nonumber
   && +\sum_i\sum_{j > i}V_{ij} \left( \hat{N}_i-1 \right)\left( \hat{N}_{j}-1 \right),
\end{eqnarray}
\end{widetext}
where the operator $\hat{c}_{i\sigma}^\dagger (\hat{c}_{i\sigma})$ creates (destroys) an electron with spin $\sigma$ in the 2p$_z$-orbital on the $i$th carbon atom. $t_i$ is the electron-transfer integral  between nearest-neighbor carbon atoms. For a chain with periodic bond alternation, we  define the single- and the double-bond transfer integrals as $t_s = t_0 (1-\delta)$ and $t_d = t_0 (1+\delta)$, respectively, where $\delta$ is a parameter which quantifies the extent of dimerization of the chain. 

The Coulomb interaction is modeled using the semi-empirical Ohno potential
\begin{equation}\label{eq:Ohno}
    V_{ij} = \frac{U}{\sqrt{1 + \left( U \epsilon_r r_{ij}/14.397 \right)^2}},
\end{equation}
where $U$ is defined as the Coulomb interaction between two charges in the same atomic orbital (in eV), $r_{ij}$ is the distance between carbon atoms $i$ and $j$ (in $\textrm{\AA}$), and $\epsilon_r$ is the relative permittivity. The number operators are defined as $\hat{N_i} = \sum_{\sigma}\hat{N}_{i\sigma}$, with $\hat{N}_{i\sigma} = \hat{c}_{i\sigma}^\dagger \hat{c}_{i\sigma}$.

We supplement the Hamiltonian given in Eqn.~\ref{eq:PPP_Ham} with an interchain term which describes the coupling between two carotenoids. We consider interchain electron hopping and Coulomb interaction between the electrons,
\begin{widetext}
\begin{eqnarray}\label{eq:interchain_ham}
    \hat{H}_{\textrm{inter}}= && -\sum_{i_1 j_2 \sigma}t^{\textrm{inter}}_{i_1j_2} \left( \hat{c}_{i_1\sigma}^\dagger \hat{c}_{j_2\sigma} + \hat{c}_{j_2\sigma}^\dagger \hat{c}_{i_1\sigma} \right)
    +\sum_{i_1j_2}V_{i_1j_2} \left( \hat{N}_{i_1}-1 \right)\left( \hat{N}_{j_2}-1 \right).
\end{eqnarray}
\end{widetext}
The appeal of this interchain coupling Hamiltonian is that it is directly comparable to the interactions described in the intrachain case, although other forms of interchain coupling have also been considered that include higher-order terms.\cite{Prodhan2017} 
In an analogous manner to the intrachain case, $t^{\textrm{inter}}_{i_1j_2}$ describes the interchain transfer integrals between carbon atom $i$ on chain 1 and carbon atom $j$ on chain 2. The Coulomb interaction $V_{i_1j_2}$ is modeled by the same Ohno potential as in Eqn.~\ref{eq:Ohno}. 

The interchain hopping matrix elements are calculated using Mulliken's approximation,\cite{Mulliken1949,Mulliken1955_2} which estimates the values of the resonance integrals $t_{ij}$ as a simple product of a constant multiplied by the overlap of the orbitals on sites $i$ and $j$. Specifically,
\begin{equation}
    t_{ij} = S_{ij} \times 9.82~\textrm{eV},
\end{equation}
where the overlap $S_{ij}$ is defined as
\begin{eqnarray}
    S_{ij} = \cos \Phi_{ij} \cos \Omega_{ij} \cos \Theta_{ij} \times S_{\pi}(r_{ij}) - \sin \Omega_{ij} \sin \Theta_{ij} \times S_{\sigma}(r_{ij}),
\end{eqnarray}
with 
\begin{equation}
    \Phi_{ij} = \arccos (\hat{\textbf{\textrm{p}}}_i \cdot \hat{\textbf{\textrm{p}}}_j),
\end{equation}
\begin{equation}
    \Omega_{ij} = \arcsin (\hat{\textbf{\textrm{p}}}_i \cdot \hat{\textbf{\textrm{r}}}_{ij})
\end{equation}
and
\begin{equation}
    \Theta_{ij} = - \arcsin (\hat{\textbf{\textrm{p}}}_j \cdot \hat{\textbf{\textrm{r}}}_{ij}).
\end{equation}
$\hat{\textbf{\textrm{p}}}_i$ and $\hat{\textbf{\textrm{p}}}_j$ are the unit vectors parallel to the p$_z$ orbitals on atoms $i$ and $j$, respectively, and $\hat{\textbf{\textrm{r}}}_{ij}$ is the unit vector parallel to the vector connecting atoms $i$ and $j$. The $\pi$ and $\sigma$ contributions to the overlap are defined as
\begin{equation}
    S_{\pi}(r) = \exp(-r\xi)\left( 1+r\xi + \frac{2}{5}(r\xi)^2 +\frac{1}{15}(r\xi)^3 \right)
\end{equation}
and 
\begin{equation}
    S_{\sigma}(r) = \exp(-r\xi)\left( -1 -r\xi -\frac{1}{5}(r\xi)^2 +\frac{2}{15}(r\xi)^3 +\frac{1}{15}(r\xi)^4 \right),
\end{equation}
where $\xi = 3.07 \textrm{\AA}^{-1}$ and $r$ is the interatomic distance.

\begin{table}[h]
    \centering
    \begin{tabular}{|w{c}{1.2 cm}|w{c}{1.2 cm}|w{c}{1.2 cm}|w{c}{1.2 cm}|}
        \hline
        $t_s$ (eV) & $t_d$ (eV) & $U$ (eV) & $\epsilon_r$ \\
        \hline
        2.2 & 2.6 & 8.0 & 2.0 \\
        \hline
    \end{tabular}
    \caption{PPP Hamiltonian parameters used in this paper.\cite{chandross_coulomb_1997}}
    \label{tab:PPP_parameters}
\end{table}

The PPP Hamiltonian parameters we use, listed in Table~\ref{tab:PPP_parameters}, are widely used to model $\pi$-conjugated systems.\cite{chandross_coulomb_1997}

The PPP Hamiltonian (Eqn.~\ref{eq:PPP_Ham}) is invariant to an electron-hole (or charge-conjugation) transformation, $\hat{c}^{\dagger}_{i\sigma}\rightarrow (-1)^i\hat{c}_{i\bar{\sigma}}$, where $\bar{\sigma}$ is the opposite spin-projection to $\sigma$.
Thus, at half-filling the eigenstates  of Eqn.~\ref{eq:PPP_Ham} are also eigenstates of the electron-hole symmetry operator $\hat{J}$, denoted by the superscripts $\pm$. In the language of valence bond theory, a $+$ ($-$) superscript denotes an ionic (covalent) state. Eqn.~\ref{eq:interchain_ham} retains this symmetry when interchain hopping is limited to specific pairs of atoms, such as between atoms that share the same $z$ coordinate (as illustrated in Fig.~\ref{fig:dimer_geometry}). We only consider this type of interchain hopping, since the exponential decay of the overlap between two orbitals with distance means that the effect of the other hopping matrix elements can be ignored.

\section{Dimer conformations}\label{ch:3}

\begin{figure}
    \centering
    \includegraphics[width=0.85\linewidth]{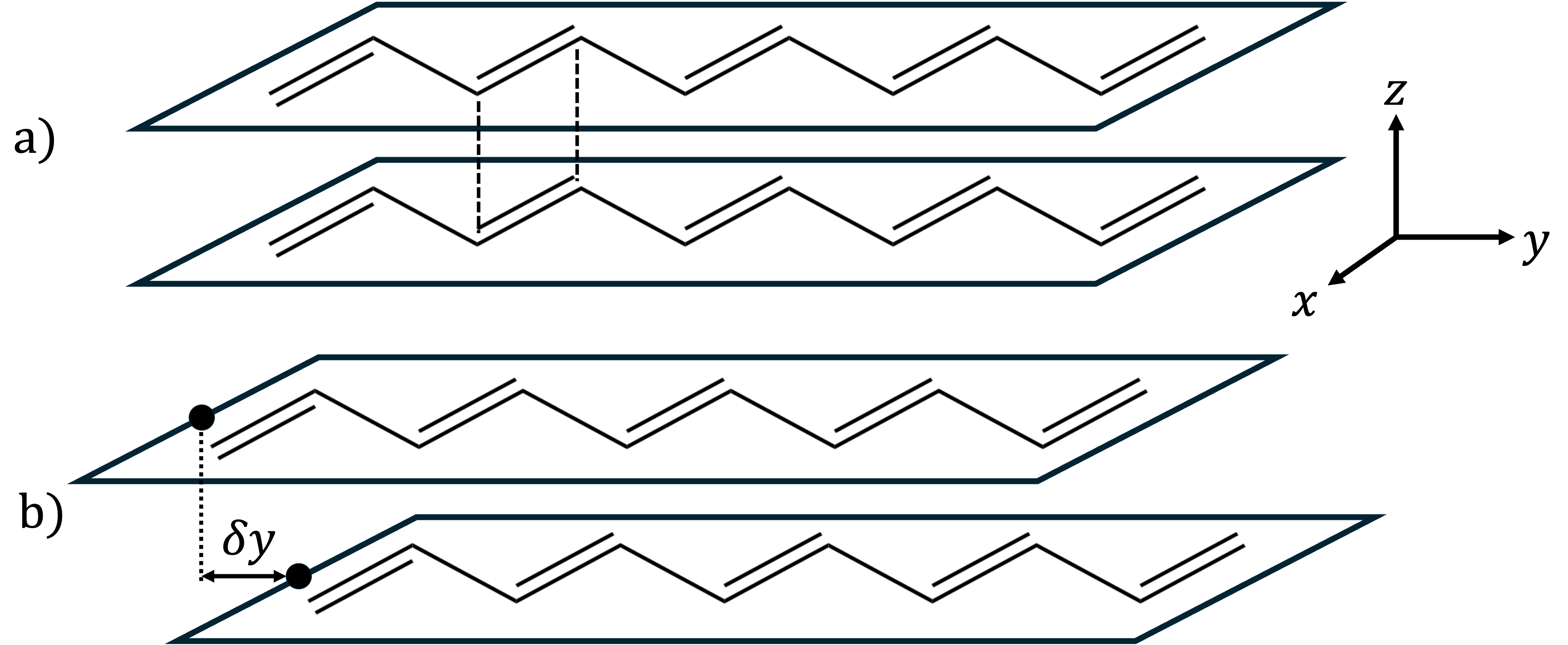}
    \caption{The arrangement of carotenoid dimers investigated in this work. In both cases, each chain contains 22 carbon atoms (which are not all shown here), representing lycopene.
    a) \textit{Vertically stacked} chains  in parallel $xy$ planes. The chains are related by a translation in the $z$ direction. This dimer belongs to the $C_{2h}$ symmetry group. b) The \textit{slip-stacked} geometry  is obtained by the translation of the lower chain along the $y$ axis, such that the first carbon atom of this chain has the same $y$ coordinate as the second carbon atom  and the same $x$ coordinate as the first carbon atom of the upper chain. This dimer belongs to the $C_{i}$ symmetry group.}
    \label{fig:dimer_geometry}
\end{figure}

The two different dimer conformations investigated in this paper are the vertically stacked dimer, illustrated in Fig.\ \ref{fig:dimer_geometry}(a), and  the slip-stacked dimer,  illustrated in Fig.\ \ref{fig:dimer_geometry}(b). 
The vertically stacked dimer has a reflection plane between the chains, and belongs to the $C_{2h}$ point group, whereas the slip-stacked dimer  has a center of inversion between the chains, and belongs to the $C_{i}$ point group.

\section{DMRG calculations}\label{ch:4}

DMRG is a method that has been used  extensively in theoretical investigations of one-dimensional systems.\cite{white_density_1992,white_density-matrix_1993} The rotation of the DMRG eigenstates from their four-block representation to the primitive-site basis using the matrix product state (MPS) tensors allows us to compute the overlap between different eigenstates. A similar procedure was employed in Ref.\cite{Ichert2026} to determine the triplet-pair weight of the $2^1\textrm{A}_\textrm{g}$ state of polyenes.  We refer the reader to Section S1 in the SI for more details of our method, and to Refs.~\cite{Schollwock2005,Schollwock2011} for reviews of the different DMRG algorithms.


In general, the eigenstates of the dimer Hamiltonian, $\hat{H} = \hat{H}_{\textrm{PPP}}+\hat{H}_{\textrm{inter}}$, can be expanded in a product basis constructed from the single-chain eigenstates $|\psi_m\rangle$ (i.e., the solutions of Eqn.~\ref{eq:PPP_Ham}):
\begin{equation}\label{eq:dimer_eig1}
|\Psi_i\rangle
=
\sum_{m,n}
\Psi_{mn}^{i}
|\psi_m^{(1)}\rangle
|\psi_n^{(2)}\rangle,
\end{equation}
where the superscripts $(1)$ and $(2)$ denote chains 1 and 2, respectively. The complete set of such product states spans the Hilbert space of the dimer. In practice, however, this basis is prohibitively large. We therefore construct a reduced dimer basis from the following low-lying single-chain eigenstates
(where the first ket represents chain 1): 
\begin{enumerate}
\item{$|1^1\textrm{A}_\textrm{g}^-\rangle|1^1\textrm{A}_\textrm{g}^-\rangle$, i.e., the two chains in their respective ground states.}
\item{$|1^1\textrm{A}_\textrm{g}^-\rangle|2^1\textrm{A}_\textrm{g}^-\rangle$ and  $|2^1\textrm{A}_\textrm{g}^-\rangle|1^1\textrm{A}_\textrm{g}^-\rangle$.}
\item{${^1}\left(|1^3\textrm{B}_\textrm{u}^-\rangle|1^3\textrm{B}_\textrm{u}^-\rangle\right) \equiv {^1}|\textrm{T}_1\textrm{T}_1\rangle$.}
\item{$|1^1\textrm{A}_\textrm{g}^-\rangle|1^1\textrm{B}_\textrm{u}^-\rangle$ and  $|1^1\textrm{B}_\textrm{u}^-\rangle|1^1\textrm{A}_\textrm{g}^-\rangle$.}
\item{$|1^1\textrm{A}_\textrm{g}^-\rangle|1^1\textrm{B}_\textrm{u}^+\rangle$ and  $|1^1\textrm{B}_\textrm{u}^+\rangle|1^1\textrm{A}_\textrm{g}^-\rangle$.}
\item{${^1}\left(|1^3\textrm{B}_\textrm{u}^-\rangle|1^3\textrm{A}_\textrm{g}^-\rangle\right) \equiv {^1}|\textrm{T}_1\textrm{T}_2\rangle$ and ${^1}|\textrm{T}_2\textrm{T}_1\rangle$.}
\item{All four lowest energy (degenerate) CT states, corresponding to an electron transfer between two chains infinitely far apart.}
\end{enumerate}

Our approach is similar in spirit to that discussed in Ref.~\cite{Young2020}, where the diabatic states (i.e., uncoupled eigenstates) are the natural building blocks of the physically relevant mixed electronic states (i.e., coupled eigenstates).

The coupled-chain eigenstate populations of the  uncoupled-chain eigenstates are $|\Psi_{mn}^{i}|^2$.
In order to simplify the discussion, we  refer to the combined population of $|1^1\textrm{A}_\textrm{g}^-\rangle|2^1\textrm{A}_\textrm{g}^-\rangle$ and  $|2^1\textrm{A}_\textrm{g}^-\rangle|1^1\textrm{A}_\textrm{g}^-\rangle$, as the total population of $|2^1\textrm{A}_\textrm{g}^-\rangle$, etc. Notice that as a consequence of using a reduced product basis, the dimer eigenstate populations displayed in Figs.~5 and 6 do not sum to unity.

We classify the eigenstates of the two dimer conformations 
according to the irreducible representations (irreps) of their symmetry groups. Table \ref{tab:irreps} shows how the dimer irreps are derived from the irreps of the single-chain eigenstates. We use DMRG to target low-energy eigenstates within specified symmetry sectors, exploiting electron-hole symmetry for both dimer geometries and, for vertically stacked dimers, reflection symmetry in the interchain plane. 

\begin{table}[htbp]
    \centering
    \renewcommand{\arraystretch}{1.3}
    \begin{tabular}{|w{c}{3.6cm}|w{c}{1.8cm}|w{c}{1.8cm}|}
        \hline
        \multirow{2}{*}{Single-chain states} & \multicolumn{2}{c|}{Dimer states} \\
        \cline{2-3}
         & VS ($C_{2h}$) & SS ($C_i$) \\
        \hline
        $\irr{A}{g}\otimes\irr{A}{g} + \irr{A}{g}\otimes\irr{A}{g}$ & $\irr{A}{g}$ & $\irr{A}{g}$ \\
        \hline
        $\irr{A}{g}\otimes\irr{A}{g} - \irr{A}{g}\otimes\irr{A}{g}$ & $\irr{A}{u}$ & $\irr{A}{u}$ \\
        \hline
        $\irr{B}{u}\otimes\irr{B}{u} + \irr{B}{u}\otimes\irr{B}{u}$ & $\irr{A}{g}$ & $\irr{A}{g}$ \\
        \hline
        $\irr{B}{u}\otimes\irr{B}{u} - \irr{B}{u}\otimes\irr{B}{u}$ & $\irr{A}{u}$ & $\irr{A}{u}$ \\
        \hline
        $\irr{A}{g}\otimes\irr{B}{u} + \irr{B}{u}\otimes\irr{A}{g}$ & $\irr{B}{u}$ & $\irr{A}{u}$ \\
        \hline
        $\irr{A}{g}\otimes\irr{B}{u} - \irr{B}{u}\otimes\irr{A}{g}$ & $\irr{B}{g}$ & $\irr{A}{g}$ \\
        \hline
    \end{tabular}
    \caption{Irreducible representations (irreps) of the dimer eigenstates derived
    from the irreps of the single-chain eigenstates. The $\irr{B}{u}$ state on
    chain~2, $|\irr{B}{u}\rangle_2$, is defined as follows. For the vertically stacked (VS) dimer, $|\irr{B}{u}\rangle_2 = \hat{\sigma}_h|\irr{B}{u}\rangle_1$,
    where $\hat{\sigma}_h$ is the reflection operator and $|\irr{B}{u}\rangle_1$ is
    the $\irr{B}{u}$ state on chain~1. For the slip-stacked (SS) dimer,
    $|\irr{B}{u}\rangle_2 = -\hat{I}|\irr{B}{u}\rangle_1$, where $\hat{I}$ is the
    inversion operator.}
    \label{tab:irreps}
\end{table}

\section{Molecular-exciton, valence-bond model of the electronic states in polyenes}\label{RM}

This section briefly introduces the molecular-exciton, valence-bond model that provides a  physically intuitive description of the excited states of polyene systems. Its primary utility here is that it enables us to derive selection rules that govern  the interchain couplings of symmetry-adapted basis states. These rules are listed in Section \ref{RM.2} and they are used in Section \ref{ch:6} to elucidate the character of the carotenoid dimer eigenstates. The next section introduces the basis states, which are described in more detail in Section S3 of the SI.

\subsection{Basis states}\label{RM.1}

The molecular-exciton, valence-bond approach, introduced by Simpson,\cite{simpson_resonance_1955} treats excitations in a conjugated polymer as being built from excitations within the monomers. The polymer linkage is then treated as a perturbation onto the set of independent simple monomer excitations. This approach was used  by Chandross \textit{et al.}\cite{Chandross1999} and Barford\cite{Barford2022} to characterize the excited states of polyenes.

\begin{figure}
    \centering
    \includegraphics[width=1\linewidth]{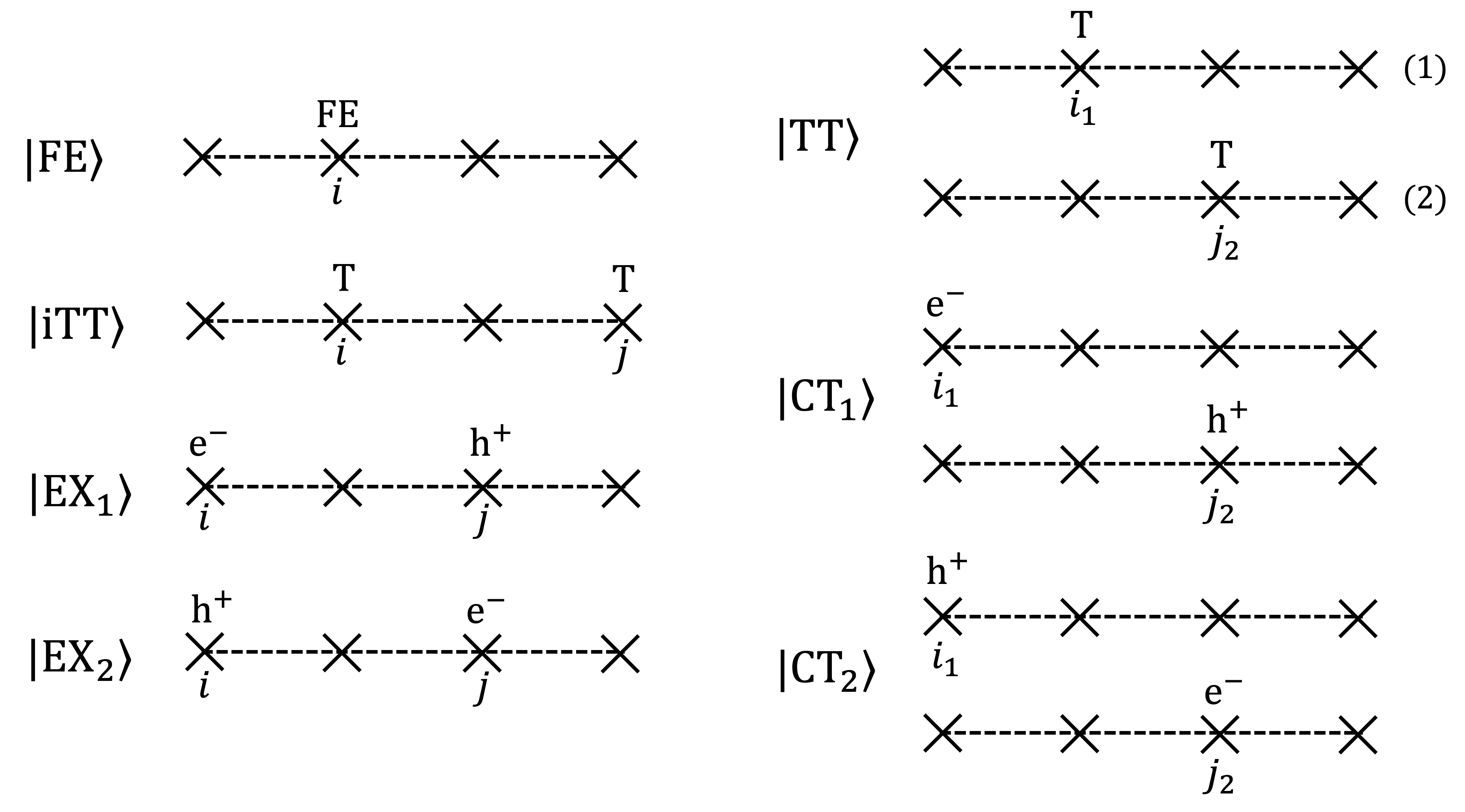}
    \caption{Schematic diagram of the basis states of the molecular-exciton, valence-bond model. The left-hand panel illustrates intrachain basis states, and the right-hand panel illustrates interchain basis states. A $\times$ symbol represents an ethylene dimer and a  dashed line represents a single bond.
    FE, EX and iTT denote the Frenkel exciton, intrachain charge-transfer exciton and intrachain triplet-pair, respectively; CT and TT denote the  interchain charge-transfer exciton and interchain triplet-pair, respectively.
    These basis states, and their spatial and electron-hole symmetry adapted counterparts, are explained in more detail in Section S3.A of the SI. 
    }
    \label{fig:RM_basis}
\end{figure}

The relevant primitive basis states are illustrated schematically in Fig.\ \ref{fig:RM_basis}. For our purposes, there are three intrachain basis states, namely the Frenkel exciton (FE), the charge-transfer exciton (EX), and the triplet-pair state (iTT). There are two interchain basis states, namely  the charge-transfer exciton (CT), and the triplet-pair state (TT). The symmetry-adapted basis states are defined in Section S3.C of the SI. Throughout this paper, we use iTT  and TT exclusively for (overall singlet) intrachain and interchain triplet-pairs, respectively.

\subsection{Selection rules for interchain couplings}\label{RM.2}



The signs of the orbital overlaps, together with  electron–hole and spatial symmetries, cause many of the matrix elements between the symmetry-adapted basis states to vanish.
The PPP model-derived  matrix elements  are listed in Sections S3.D and S3.E of the SI.
These results lead to the following selection rules governing interchain coupling:

\begin{enumerate}
    \item{For vertically stacked dimers:}
    \begin{enumerate}
        \item{Out-of-phase (i.e., dark) intrachain excitons (composed of the FE and EX basis states) couple to out-of-phase interchain charge-transfer excitons  (composed of the  CT  basis states); see Eqns.~S34 and S35.
        } 
        \item{Conversely, in-phase (i.e., bright) intrachain excitons are not coupled to any interchain state.}
        \item{There is no coupling between interchain charge-transfer excitons (CT) and interchain triplet-pair states (TT).}
        \item{In-phase intra- and interchain triplet-pair (iTT and TT, respectively) states are coupled;  see Eqn.~S36.}
    \end{enumerate}
    \item{For slip-stacked dimers:}
        \begin{enumerate}
        \item{In contrast to vertically stacked dimers,  in-phase (i.e., bright) intrachain excitons (linear combinations of FE and $\textrm{EX}$) couple to in-phase interchain charge-transfer (CT) excitons; see Eqns.~S37 --  S40.} 
        \item{Out-of-phase (i.e., dark) intrachain excitons are not coupled to any interchain state.}
        \item{There is coupling between the interchain charge-transfer (CT) excitons and interchain triplet-pair (TT) states; see Eqn.~S41.}
        \item{In-phase intra- and interchain triplet-pair (iTT and TT, respectively) states are coupled; see Eqn.~S42.}
    \end{enumerate}
\end{enumerate}

\begin{figure}[htbp]
\centering
\includegraphics[width=0.8\linewidth]{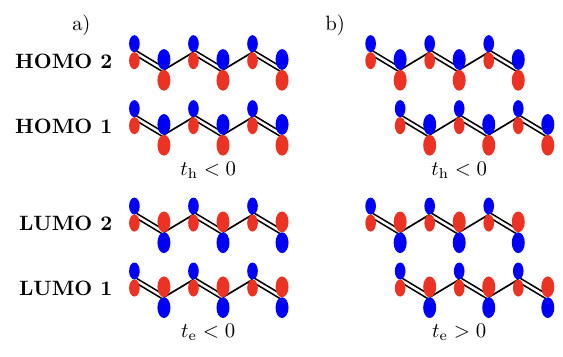}
\caption{The signs of the interchain electron and hole-transfer integrals ($t_\textrm{e}$ and $t_\textrm{h}$, respectively) for a) vertical- and b) slip-stacked dimers. The color coding of the atomic-orbital represents their relative signs in the HOMOs and LUMOs.
The difference in the relative signs of the electron and hole-transfer integrals accounts for the different selection rules governing interchain coupling.}
\label{hole_transfer}
\end{figure}

These selection rules  imply a qualitative distinction between the two conformations. 
In a vertically stacked dimer, coupling between interchain CT excitons and both the bright-dimer state and interchain triplet-pairs vanishes. However, both couplings are allowed for slip-stacked dimers. This difference in couplings arises from the  opposite signs of the interchain hole-transfer integrals, as illustrated in Fig.\ \ref{hole_transfer}.

The selection rules help us to interpret the different wavefunction components for vertical and slip-stacked dimers, as  described in Section \ref{ch:6}. In addition, as discussed in Section \ref{ch:7}, this difference between the couplings  prevents a possible CT-mediated SF mechanism in vertically stacked dimers, but allows it in slip-stacked dimers.

\section{Results}\label{ch:6}

\subsection{Lycopene Monomers}\label{ch:6.1}

\begin{table}[h]
    \centering
    \begin{tabular}{|c|c|}
        \hline
        State & Energy (eV) \\
        \hline
        $\mathrm{T}_1$ ($1^3\mathrm{B}_{\mathrm{u}}^{-}$) & 1.37 \\
        \hline
        $\mathrm{T}_2$ ($1^3\mathrm{A}_{\mathrm{g}}^{-}$) & 1.68 \\
        \hline
        $2^1\mathrm{A}_{\mathrm{g}}^{-}$ & 2.52 \\
        \hline
        $1^1\mathrm{B}_{\mathrm{u}}^{-}$ & 2.86 \\
        \hline
        $1^1\mathrm{B}_{\mathrm{u}}^{+}$ & 3.00 \\
        \hline
        CT($r_\textrm{inter}\rightarrow\infty$) & 4.28\\
        \hline
    \end{tabular}
    \caption{PPP model calculated vertical excitation energies of the single-chain eigenstates of lycopene. Also shown is the energy of a pair of anionic and cationic lycopene chains infinitely far apart.}
    \label{table:single_chain}
\end{table}

The excitation energies of the single-chain eigenstates considered in this paper are listed in Table~\ref{table:single_chain}. The two lowest-energy eigenstates are the triplets, $\mathrm{T}_1$ and $\mathrm{T}_2$. For the PPP parameters used here, the lowest-energy singlet eigenstate is the highly-correlated, dark $2^1\mathrm{A}_{\mathrm{g}}^{-}$ state. As is well-documented,\cite{Chandross1999,Barford2022} this state has both  triplet-pair  and  charge-transfer exciton character. In terms of the basis states introduced in Section \ref{RM.1}  and in Section S3.A of the SI, it is of the form,
\begin{equation}\label{eq:dimer_eig}
    |2^1\mathrm{A}_{\mathrm{g}}^{-}\rangle = \sum_{ij} a_{ij} |\mathrm{iTT;i,j}\rangle +
    b_{ij} |\mathrm{EX^-;i,j}\rangle,
\end{equation}
where $|\mathrm{iTT;i,j}\rangle$ and $|\mathrm{EX^-;i,j}\rangle$ are defined in Eqns.~S8 and S21, respectively. The triplet-pair population of the $2^1\mathrm{A}_{\mathrm{g}}^{-}$ state (i.e., $\sum_{ij}|a_{ij}|^2$) has been estimated at ca.\ 75\%.\cite{Starcke2006,Boguslawski2019,Ichert2026} As shown in Ref.\cite{Barford2022}, the hybridization of the iTT and EX components of the $2^1\mathrm{A}_{\mathrm{g}}^{-}$ state causes a strong triplet-pair attraction. This intrachain triplet-pair attraction means that the $2^1\mathrm{A}_{\mathrm{g}}^{-}$ state has a lower energy than the  interchain triplet-pair state $\textrm{T}_1\textrm{T}_1$ in  an uncoupled dimer, as illustrated in Fig.\ \ref{fig:uncoupled}. The $1^1\mathrm{B}_{\mathrm{u}}^{-}$ state is another predominantly triplet-pair state, being a higher-momentum replica of the $2^1\mathrm{A}_{\mathrm{g}}^{-}$ state.\cite{Valentine2020}

The next singlet eigenstate is the bright, predominantly Frenkel exciton state
\begin{equation}\label{eq:1Bu}
    |1^1\mathrm{B}_{\mathrm{u}}^+\rangle = \sum_{i} c_{i} |\mathrm{FE;i}\rangle +
   \sum_{ij} d_{ij} |\mathrm{EX^+;i,j}\rangle,
\end{equation}
where $|\mathrm{FE;i}\rangle$   and $|\mathrm{EX^+;i,j}\rangle$ are defined in Eqns.~S6 and S21, respectively. Eqn.~\ref{eq:1Bu} indicates that the $1^1\mathrm{B}_{\mathrm{u}}^{+}$ state also contains some even electron-hole parity intrachain charge-transfer exciton character.

\subsection{Uncoupled Lycopene Dimers}\label{ch:6.2}

\begin{figure}
     \centering
     \includegraphics[width=0.9\linewidth]{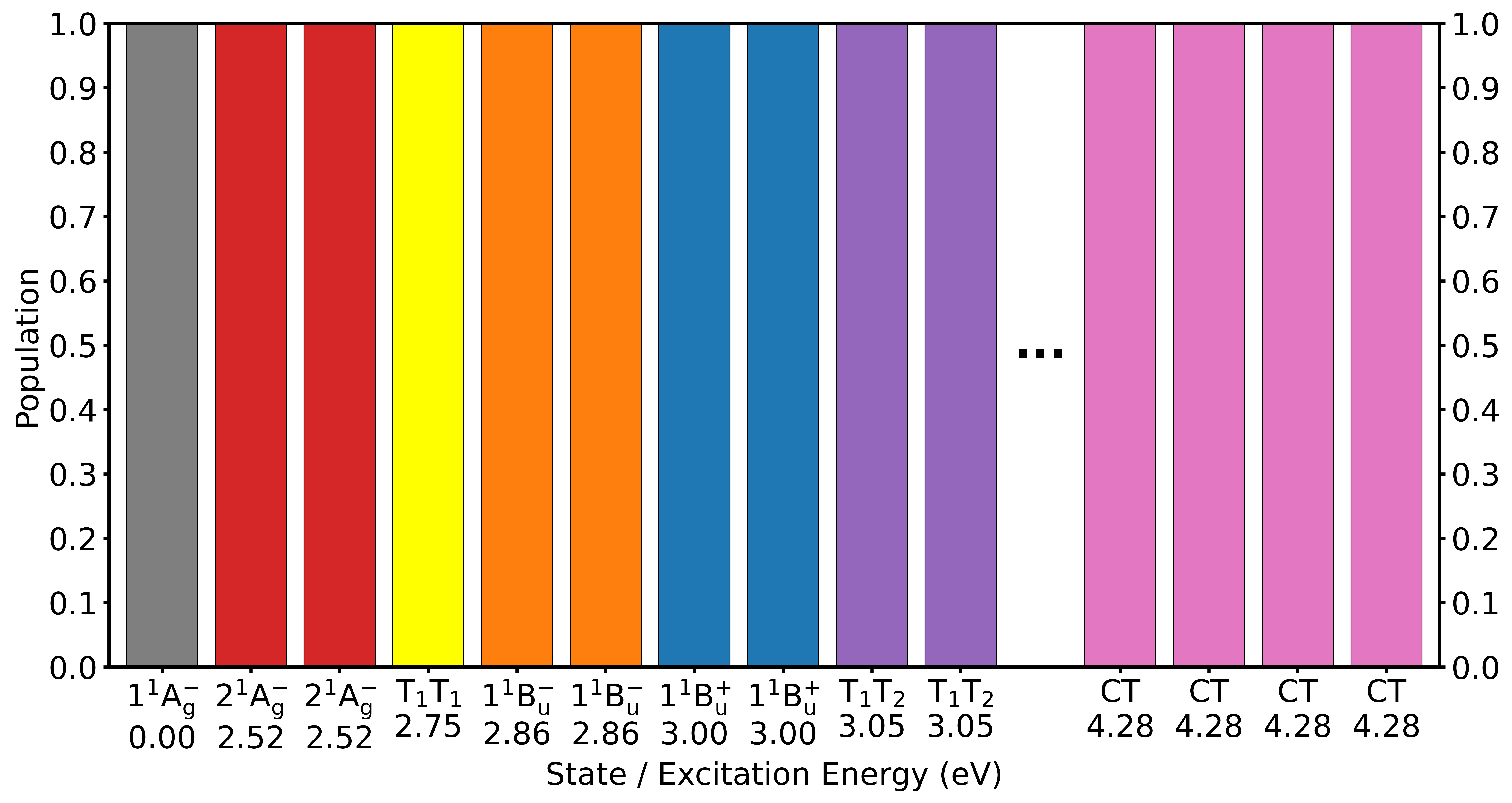}
     \caption{Eigenstates and excitation energies (in eV) of the uncoupled lycopene dimer. These are identified by their single-chain symmetry labels.}
     \label{fig:uncoupled}
 \end{figure}

Before discussing the coupled-dimer eigenstates, we describe the uncoupled-dimer eigenstates, obtained directly from the single-chain states described in Section \ref{ch:6.1}.
Fig.\ \ref {fig:uncoupled} illustrates the energies of the eigenstates for an uncoupled dimer in ascending value, which mirror the list given in Section \ref{ch:4}. As already noted, as a consequence of the intrachain triplet-pair attraction, for uncoupled chains the \textit{intrachain} triplet-pair state,  $2^1\mathrm{A}_{\mathrm{g}}^{-}$, lies lower in energy than the \textit{interchain} triplet-pair state, $\textrm{T}_1\textrm{T}_1 \equiv 1^3\mathrm{B}_{\mathrm{u}}^{-} \otimes1^3\mathrm{B}_{\mathrm{u}}^{-}$.  $\textrm{T}_1\textrm{T}_2 \equiv 1^3\mathrm{B}_{\mathrm{u}}^{-} \otimes1^3\mathrm{A}_{\mathrm{g}}^{-}$ are higher-energy interchain triplet-pair states. CT are the (uncoupled) interchain charge-transfer states. 

Note that, within the reduced model, the uncoupled interchain triplet-pairs are approximated by
\begin{equation}
    {^1}|\textrm{T}_m \textrm{T}_n\rangle = \sum_{ij}f_{ij}^{mn}|\textrm{T}\textrm{T};i,j\rangle,
\end{equation}
where $m$ and $n$ are the excitation number of the single-chain triplet states, and $|\textrm{T}\textrm{T};i,j\rangle$ are the interchain reduced model basis states, defined in Eqn.~S10. A triplet-pair can have a singlet, triplet or quintet spin multiplicity, but as we are only concerned with singlet triplet-pairs in this paper, we drop the multiplicity superscript.

\subsection{Vertically Stacked Lycopene Dimers}\label{ch:6.3}

We begin the discussion of our results with the spectra of a vertically stacked lycopene dimer. The  interchain separations, $r_\textrm{inter}$, are 3.0 \AA~and 3.5 \AA, corresponding to interchain transfer integrals of $t_{\textrm{inter}}$ of $ 0.55$ eV and $0.22$ eV, respectively. These interchain distances have been previously used in theoretical calculations,\cite{Aryanpour2015,Prodhan2017,Barford2023} and can be associated with strongly and weakly coupled chains, respectively. 
The stronger-coupling case also illustrates the enhanced stabilization of interchain configurations expected as the effective intermolecular interaction increases, for example, via multiple neighbors in an aggregate.

\begin{figure}[]
    \centering
    \begin{subfigure}[t]{0.80\textwidth}
        \centering
        \caption{}
        \includegraphics[width=\textwidth]{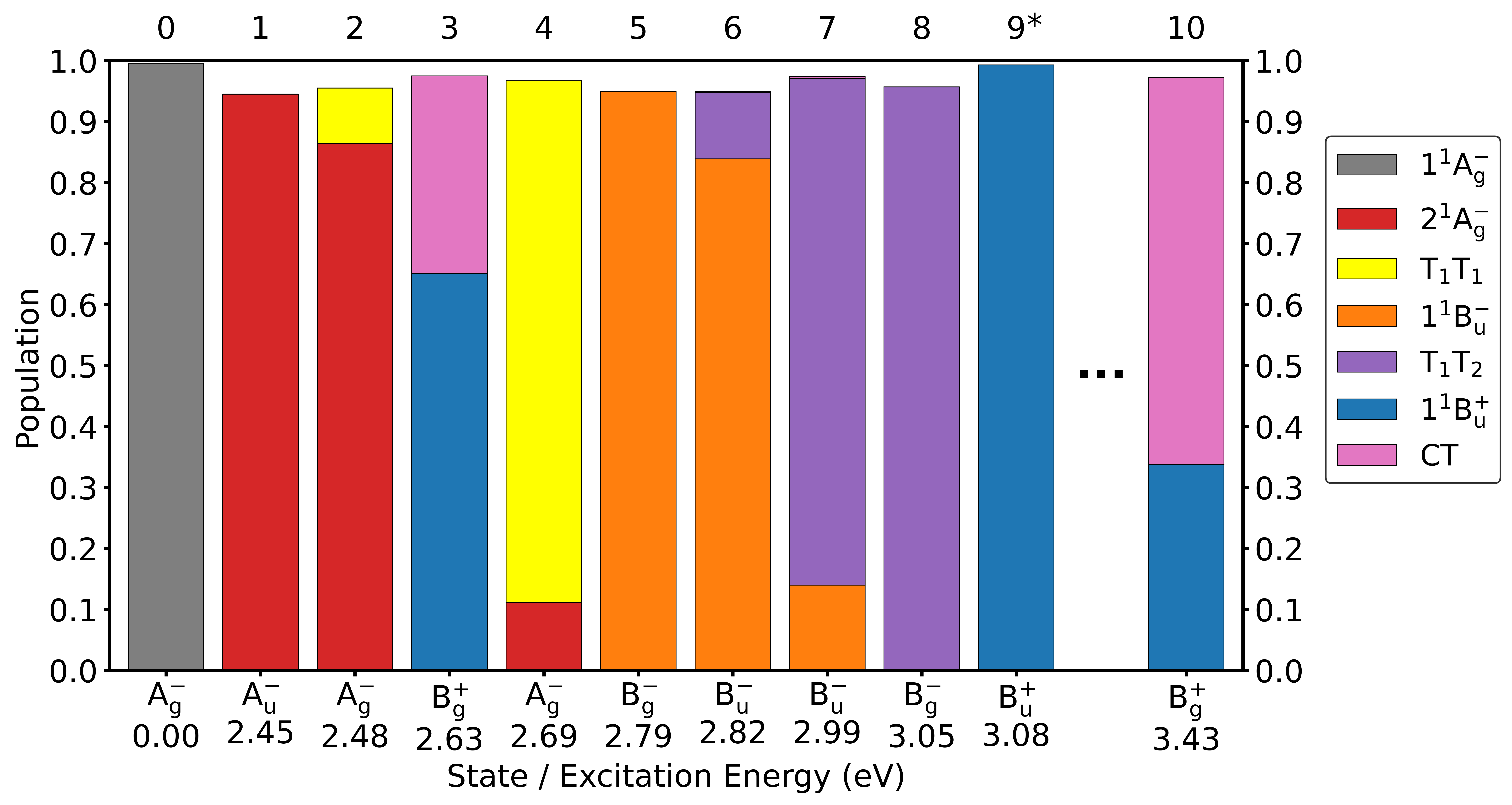}
        \label{fig:VS_35}
    \end{subfigure}
    \hfill
    \begin{subfigure}[t]{0.85\textwidth}
        \centering
        \caption{}
        \includegraphics[width=\textwidth]{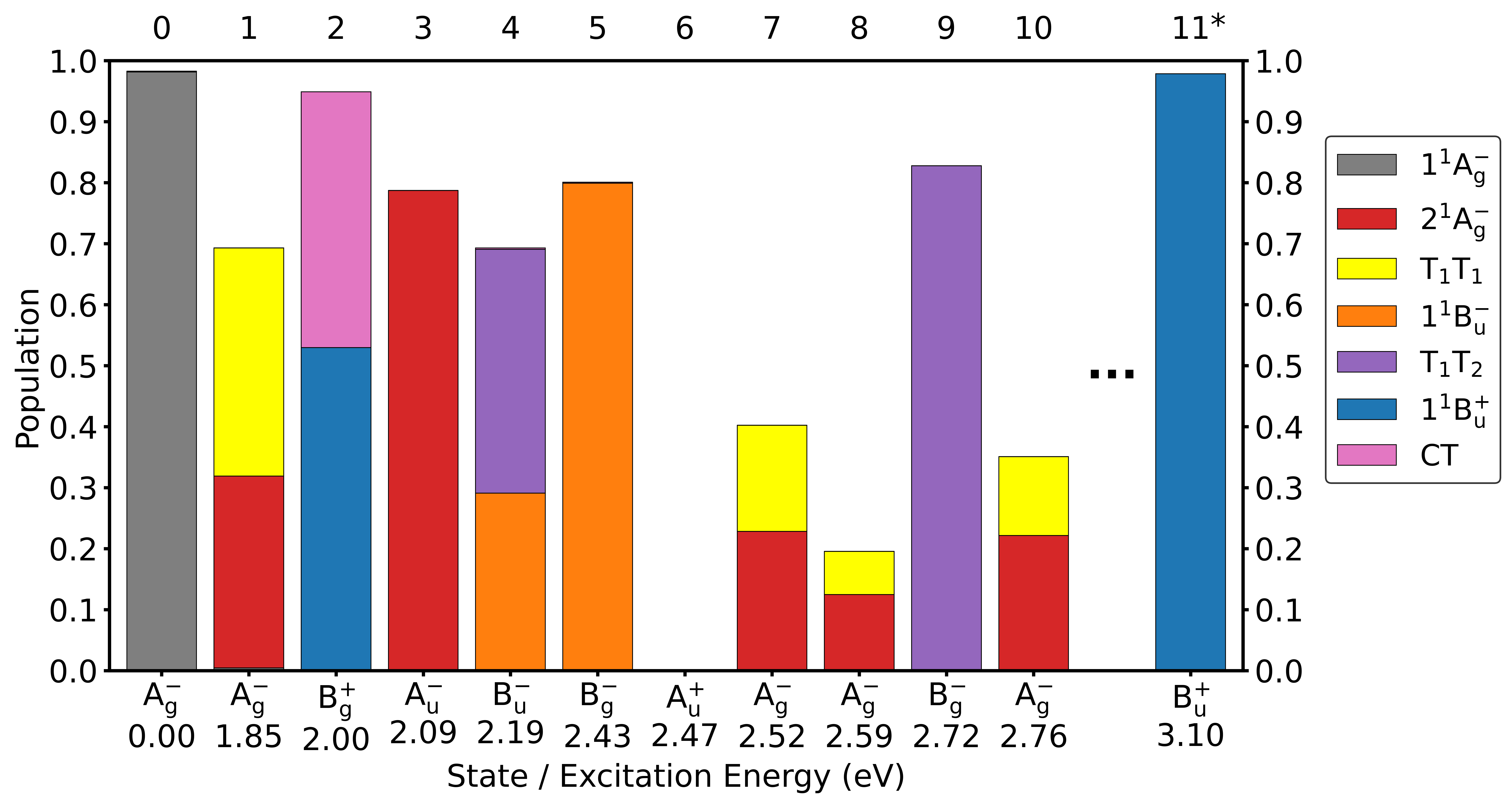}
        \label{fig:VS_30}
    \end{subfigure}
    \begin{small}
    \caption{State populations in terms of the single-chain eigenstates (shown in the panel), symmetry labels (irreps) and excitation energies of the eigenstates for the vertically stacked dimer.  
    The irreps of the dimer eigenstates are determined from the irreps of the single-chain eigenstates, as shown in Table \ref{tab:irreps}.
    * indicates the bright-dimer state.
    Interchain distances of (a) 3.5 \AA~ (i.e., $t_{\textrm{inter}}=0.22$ eV) and (b) 3.0 \AA~(i.e., $t_{\textrm{inter}}= 0.55$ eV).
    The ellipses represent higher-energy dimer states within particular symmetry sectors that are computationally prohibitive to calculate.}
    \end{small}
    \label{fig:vertically_stacked}
\end{figure}

Fig.\ 5(a) shows the spectrum at $r_\textrm{inter} = 3.5~\textrm{\AA}$. The character of these states can be readily understood from the selection rules introduced in Section \ref{RM.2}.
\begin{enumerate}
    \item{
    The bright-dimer state ($\textrm{B}_\textrm{u}^+$, No.\ 9), identified by its non-zero transition dipole moment  with the ground state,  has an excitation energy of 3.08 eV. Alternatively, based on  symmetry and the projection of this state onto the uncoupled chain eigenstates, we identify this state as the in-phase linear combination of predominantly Frenkel excitons (FE) (with some \textit{intrachain} charge-transfer (EX) components). Importantly, as predicted by the selection rule 1(b), this state has no \textit{interchain} charge-transfer exciton components.}
 \item{The out-of-phase FE combination ($\textrm{B}_\textrm{g}^+$, No.\ 3), i.e., the dark FE state, is  at 2.63 eV. As predicted by selection rule 1(a), this state does have  interchain charge-transfer character. Its complementary state (No.\ 10) has predominantly CT character and is also optically inaccessible.}
 \item{
 The negative electron-hole symmetry states 1, 2 and 4 form a trio of predominantly triplet-pair states, being   linear combinations of the intrachain $2^1\textrm{A}_\textrm{g}^-$ and interchain $\textrm{T}_1\textrm{T}_1$ states. The amount of interchain triplet-pair character increases with energy. 
 As predicted by selection rule 1(d), the in-phase linear combinations (No.~2 and 4) have mixed intra- and interchain triplet-pair character, whereas the out-of-phase combination (No.~1) has only intrachain triplet-pair character. Selection rule 1(c) predicts that, to first-order, the interchain $\textrm{T}_1\textrm{T}_1$ state cannot couple to a CT state.
 However, Eqn.~S36 of the SI indicates that interchain triplet-pair coupling is a second-order process mediated by a virtual CT exciton. In addition, in principle, the intrachain charge-transfer exciton component of the $2^1\textrm{A}_\textrm{g}^-$ state (i.e., EX) can couple to interchain charge-transfer components, but again  these are virtual processes involving high-energy charge-transfer states of $\textrm{A}_\textrm{g}^-$ symmetry, which are excluded from our reduced product basis.
}
 \item{States 5-8 form a quartet of predominantly triplet-pair states, being linear combinations of intrachain $1^1\textrm{B}_\textrm{u}^-$, and interchain $\textrm{T}_2\textrm{T}_1$ and $\textrm{T}_1\textrm{T}_2$ states. The out-of-phase combinations 5 and 8 have either pure intra- or interchain triplet-pair character, respectively. }
\end{enumerate}

The spectrum for the same dimer geometry at $r_\textrm{inter} = 3.0$ \AA~is shown in Fig.\ 5(b). An analogous analysis as above shows that this spectrum follows the same selection rules (and symmetry arguments) for the mixing of uncoupled states. However, as a consequence of stronger interchain coupling, there are some important changes to the energetic ordering and character of the eigenstates. 

\begin{enumerate}
    \item{Stronger interchain coupling increases the contributions of higher-energy single-chain states to most of the dimer excited states. 
    }
    \item{Significantly, stronger interchain coupling results in a reordering of the intrachain $2^1\textrm{A}_\textrm{g}^-$ and interchain $\textrm{T}_1\textrm{T}_1$ combinations (1 and 3 in Fig.\ 5(b)), with the lower state  having slightly more interchain than intrachain triplet-pair character. }
    \item{A similar reordering is observed for the $\textrm{T}_1\textrm{T}_2$, $\textrm{T}_2\textrm{T}_1$ and $1^1\textrm{B}_\textrm{u}^-$ linear combinations, with again the lowest state (4) of the quartet having more interchain than intrachain triplet-pair character.}
    \item{Based on its symmetry, the  $\textrm{A}_\textrm{u}^+$ state (No.~6) is assumed to have contributions from the single-chain $1^1\textrm{A}_\textrm{g}^+$ state (i.e., the second lowest energy FE state), which is excluded from our reduced product basis.}

\end{enumerate}

\subsection{Slip-Stacked Lycopene Dimers}\label{sect6.4}

\begin{figure}[htbp]
    \centering
    \begin{subfigure}[t]{0.80\textwidth}
        \centering
        \caption{}
        \includegraphics[width=\textwidth]{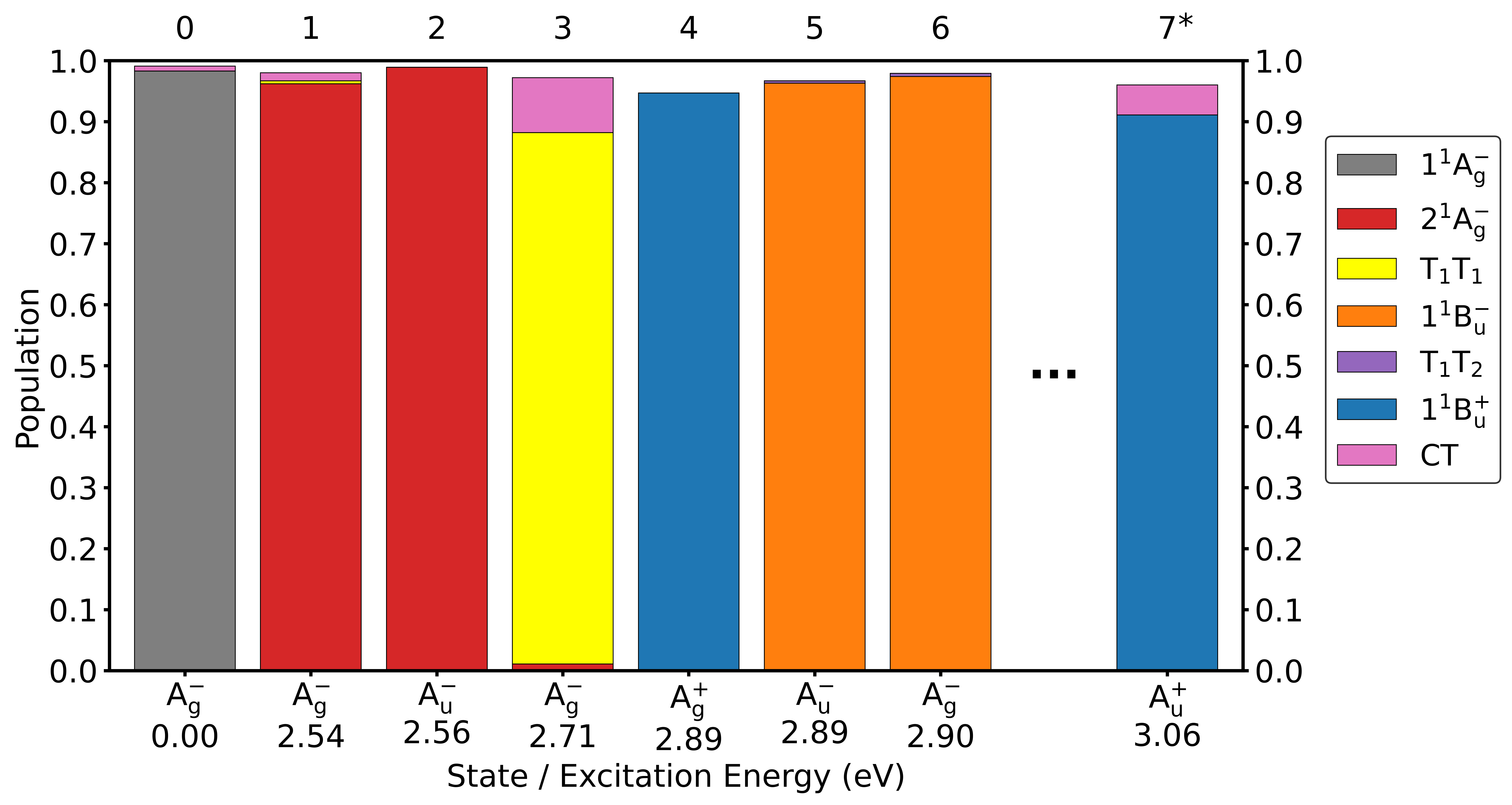}
        \label{fig:SS_35}
    \end{subfigure}
    \hfill
    \begin{subfigure}[t]{0.85\textwidth}
        \centering
        \caption{}
        \includegraphics[width=\textwidth]{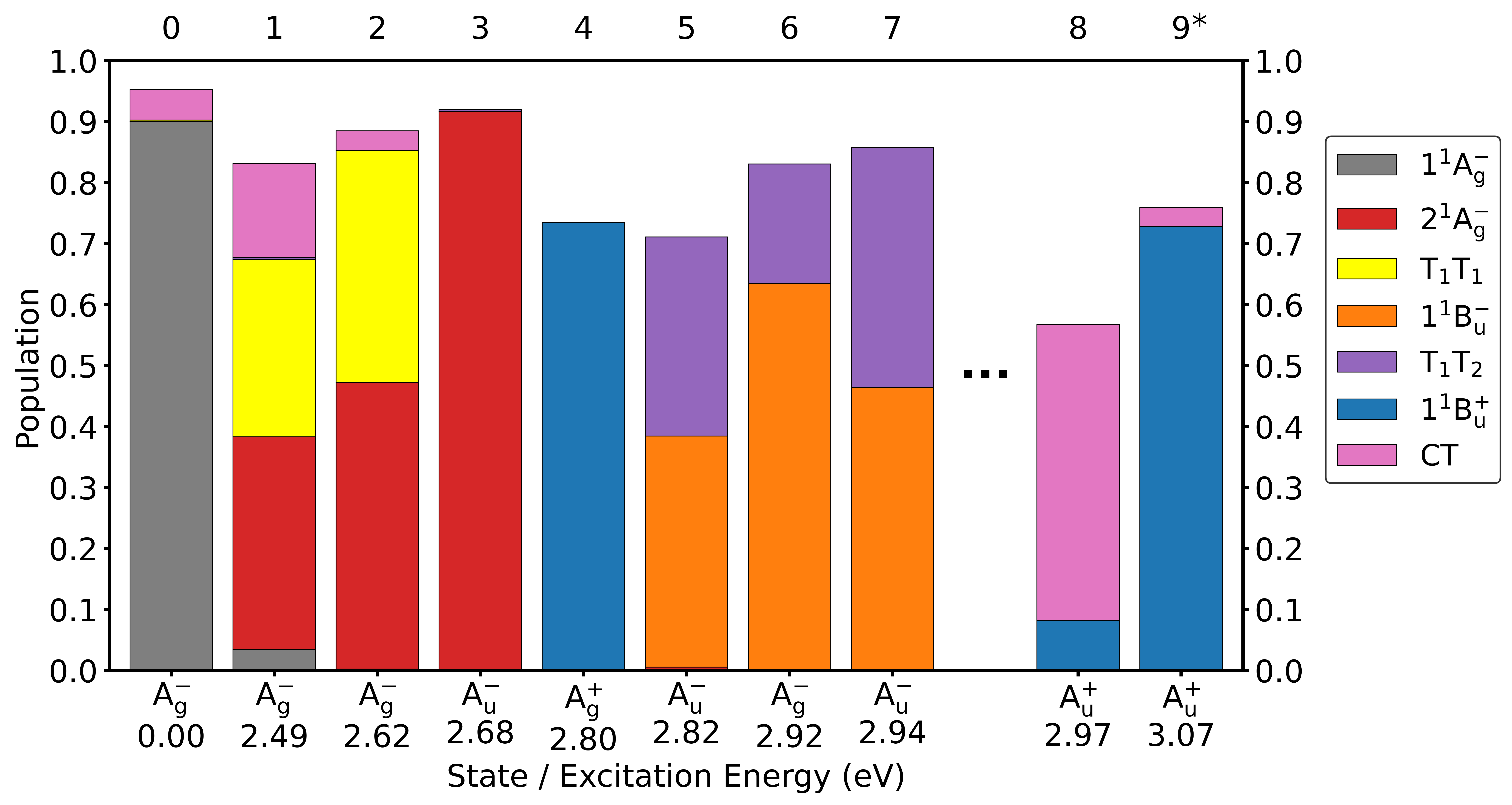}
        \label{fig:SS_30}
    \end{subfigure}
    \begin{small}
    \caption{State populations in terms of the single-chain eigenstates (shown in the panel), symmetry labels (irreps) and excitation energies of the eigenstates for the slip-stacked dimer.
        The irreps of the dimer eigenstates are determined from the irreps of the single-chain eigenstates, as shown in Table \ref{tab:irreps}.
     * indicates the bright-dimer state.
    Interchain distances of (a) 3.5 \AA~(i.e., $t_{\textrm{inter}}= 0.18$ eV)  and (b) 3.0~\AA~(i.e., $t_{\textrm{inter}}= 0.45$ eV). 
    The ellipses represent higher-energy dimer states within particular symmetry sectors that are computationally prohibitive to calculate.
    }
    \end{small}
    \label{fig:slip_stacked}
\end{figure}

We now describe the spectra of a slip-stacked dimer shown in Fig.\ 6
for chain separations of 3.0 and 3.5 \AA, corresponding to a $t_{\textrm{inter}}$
of  0.45 eV and 0.18 eV, respectively.

Unlike the previous case of perfectly aligned, vertically stacked molecules, the chains are no longer related by an intermolecular plane of symmetry. The misalignment of the molecules has a significant impact on the coupling of the uncoupled states.

\begin{enumerate}
\item{The number of eigenstates with contributions  from interchain CT excitons  has increased. In the slip-stacked arrangement, as predicted by the selection rules 2(a) and 2(b), CT states  no longer couple to the out-of-phase (dark) FE combination, but instead couple to the in-phase  (bright) FE (i.e., No.~7 in Fig.\ 6(a) and No.~8 and 9 in Fig.\ 6(b)). As a consequence of this mixing, we identify two (complementary) dipole-allowed states in Fig.\ 6(b), i.e., 8 and 9.
Because of its much larger CT character, state 8 has an oscillator strength ten times smaller than state 9; an observation which has implications for a CT-mediated SF pathway, as discussed in Section \ref{ch:7}.}

\item{In contrast to vertically stacked dimers, and as predicted by selection rule 2(c), the linear combinations of $2^1\textrm{A}_\textrm{g}^-$ and $\textrm{T}_1\textrm{T}_1$ now have interchain charge-transfer  (CT) character.}

\item{Decreasing (increasing) the chain separation (coupling) causes a mixing of states 1 and 3, and a reversal in energy of states 2 and 3 of Fig.\ 6(a), resulting in an increase in both the interchain triplet-pair and charge-transfer exciton character of state 1, as shown in Fig.~6(b).}

\item{As in vertically stacked dimers, stronger interchain coupling increases the contributions of higher-energy single-chain states to the low-energy dimer eigenstates. }
\end{enumerate}

\section{Discussion and Conclusions} \label{ch:7}

In this paper, we have characterized the low-lying eigenstates of  lycopene dimers comprising two 22-carbon-atom chains. Using numerically accurate DMRG solutions of the Pariser-Parr-Pople model, we determined the populations of the dimer eigenstates in terms  of the eigenstates of the individual chains. Complementing these calculations, we employed a molecular-exciton, valence-bond model of polyene systems to derive  selection rules governing interchain coupling, which provide a physical framework for interpreting the DMRG results.

Our calculations indicate two distinct effects of intermolecular coupling. First, the dimer geometry determines which intrachain and interchain basis states couple. Second, increasing the strength of the interchain coupling  preferentially stabilizes states with greater interchain character, leading in some cases to a reordering of the low-lying eigenstates. Together, these effects determine which electronic-state pathways are energetically accessible for singlet fission.

\begin{figure}[htbp]
\centering
\includegraphics[width=8 cm]{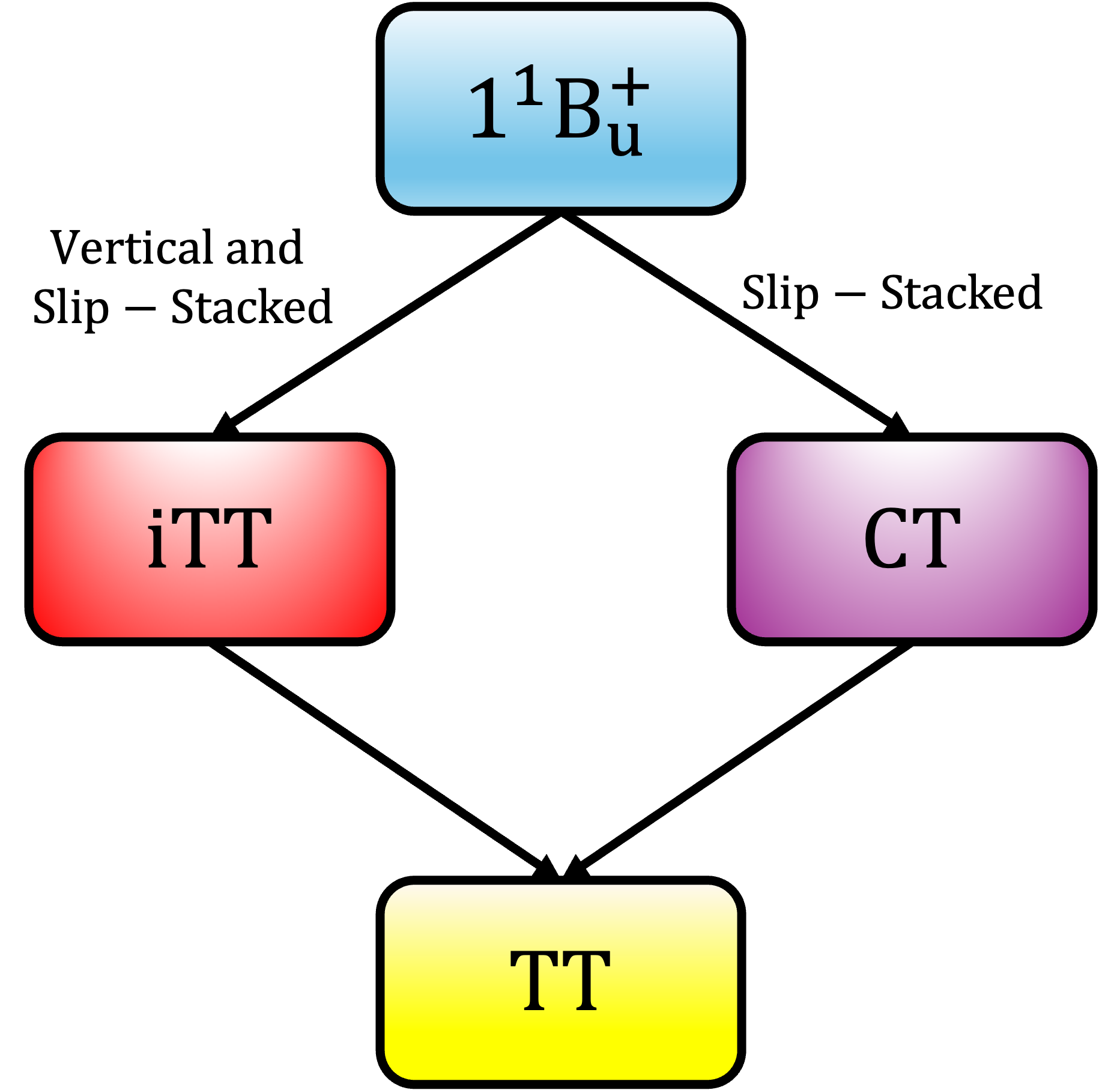}
\caption{Energetically viable SF pathways for vertically stacked and slip-stacked carotenoid dimers. \buplus{1} denotes the bright-dimer state and TT the interchain triplet-pair state. The possible intermediate states are an intrachain triplet-pair state (iTT) and an interchain charge-transfer exciton (CT).}
\label{fig7}
\end{figure}

For a vertically stacked dimer, the  selection rules prevent both the bright-dimer state and the low-lying triplet-pair states from acquiring interchain CT character. In the weak-coupling regime, the lowest-energy excited state has purely intrachain character and is predominantly the triplet-pair \agminus{2} state. As the interchain coupling increases, the energetic ordering of the triplet-pair states is reversed, and the lowest-energy state acquires increasing interchain triplet-pair character. Thus, although stronger coupling favors the formation of an interchain triplet-pair, no low-lying CT intermediate connects the photoexcited bright state to the triplet-pair manifold. Within the states considered here, SF must therefore proceed through an intermediate intrachain triplet-pair state. Using the state labels of Fig.~5(b), the corresponding energetically viable pathways are $11\rightarrow3\rightarrow1$ (i.e., via the \agminus{2} state) and $11\rightarrow5\rightarrow4$ (i.e., via the \buminus{1} state) as illustrated schematically in Fig.~\ref{fig7}.

The situation is qualitatively different for a slip-stacked dimer, where coupling of both the bright-dimer and interchain triplet-pair states to interchain CT excitons is allowed. At stronger interchain coupling, the lowest-energy eigenstate consequently develops substantial contributions from both interchain triplet-pair and CT configurations. The bright-dimer state carrying the largest oscillator strength has only a small CT contribution, whereas a lower-energy, weakly dipole-allowed state acquires substantial CT character. This provides an additional energetically accessible CT-mediated pathway to the interchain triplet-pair manifold. Using the state labels of Fig.~6(b), this pathway is $9\rightarrow8\rightarrow1$, as illustrated in Fig.~\ref{fig7}. An iTT-mediated pathway, such as $9\rightarrow3\rightarrow1$, also remains energetically accessible. Our calculations suggest, however, that because  of its weak oscillator strength, direct formation of the interchain triplet-pair following photoexcitation into a predominantly CT exciton is unlikely.

These two dimer conformations  illustrate the roles of geometry and coupling strength in determining the accessible SF pathways. Geometry determines which electronic configurations can mix, whereas the magnitude of the interchain coupling controls their degree of mixing and energetic ordering. In particular, increasing the interchain coupling  stabilizes interchain triplet-pair configurations relative to their intrachain counterparts and, for slip-stacked dimers, stabilizes CT-rich states below the predominantly bright-dimer state. We therefore reach conclusions consistent with those of Santra \textit{et al.}\cite{Santra2022,Santra2025}: both intrachain-triplet-pair-mediated and CT-mediated SF pathways are possible in carotenoid dimers, with molecular packing determining which pathways are allowed and energetically accessible.

The two dimer geometries considered here represent idealized limiting cases of the molecular arrangements that may occur in carotenoid aggregates. In particular, the selection rules for the vertically stacked geometry arise from its higher intermolecular symmetry, and thus distortions that break this symmetry will allow otherwise forbidden couplings. Structural disorder in intermolecular separation, slip, and relative orientation is therefore expected to continuously tune both the magnitude of the electronic couplings and the energetic ordering of the resulting states. The vertically stacked and slip-stacked geometries should be regarded  as limiting cases that illustrate the  relationship between molecular packing and excited-state character. Disorder in realistic aggregates may allow multiple SF pathways to coexist.

The SF pathways identified in this paper are determined from the character, symmetry, and energetic ordering of the electronic eigenstates, rather than from an explicit treatment of excited-state dynamics. Computing their relative efficiencies  requires a dynamical description. Moreover, although the dimer provides a natural starting point for understanding these processes, the excited-state structure of an extended aggregate may differ substantially from that of an isolated dimer.\cite{Aryanpour2015}
A promising protocol for addressing both issues is to construct a coarse-grained description of carotenoid aggregates parametrized by monomer and dimer electronic-structure calculations. Such an approach was recently used by Santra \textit{et al.}\cite{Santra2026} for acene aggregates. For our purposes, the dimer eigenstates can be used to determine effective coupling matrix elements between low-lying states of neighboring carotenoids, providing a tractable model for larger aggregates. Such a model would provide a natural framework for future studies of excited-state dynamics and the relative efficiencies of competing SF pathways in carotenoid aggregates.


\begin{acknowledgements}
A.G.I. is grateful for financial support from the Department of Chemistry, University of Oxford;
and Linacre College for the Carolyn and Franco Gianturco Scholarship.
The authors  acknowledge the use of the University of Oxford Advanced Research Computing facility in carrying out this work. 
\end{acknowledgements}

\section*{Supporting Information}
The Supporting Information contains further details of the DMRG method, MPS wavefunction mapping, and the derivation of the interchain coupling selection rules determined via the molecular-exciton, valence-bond method.

\vfill\pagebreak

\bibliography{AI_refs.bib}

@article{marcus_triplet-triplet_2020,
	volume = {102},
	issn = {2469-9950, 2469-9969},
	url = {https://link.aps.org/doi/10.1103/PhysRevB.102.035134},
	doi = {10.1103/PhysRevB.102.035134},
	number = {3},
	urldate = {2025-10-06},
	journal = {Physical Review B},
	author = {Marcus, Max and Barford, William},
	month = jul,
	year = {2020},
	pages = {035134},
}

@article{khokhlov_ab_2020,
	volume = {124},
	copyright = {https://doi.org/10.15223/policy-029},
	issn = {1089-5639, 1520-5215},
	url = {https://pubs.acs.org/doi/10.1021/acs.jpca.0c01678},
	doi = {10.1021/acs.jpca.0c01678},
	language = {en},
	number = {28},
	urldate = {2026-04-12},
	journal = {The Journal of Physical Chemistry A},
	author = {Khokhlov, Daniil and Belov, Aleksandr},
	month = jul,
	year = {2020},
	pages = {5790--5803},
}

@article{chandross_coulomb_1997,
	volume = {55},
	copyright = {http://link.aps.org/licenses/aps-default-license},
	issn = {0163-1829, 1095-3795},
	url = {https://link.aps.org/doi/10.1103/PhysRevB.55.1497},
	doi = {10.1103/PhysRevB.55.1497},
	language = {en},
	number = {3},
	urldate = {2026-04-21},
	journal = {Physical Review B},
	author = {Chandross, M. and Mazumdar, S.},
	month = jan,
	year = {1997},
	pages = {1497--1504},
}

@article{white_density-matrix_1993,
	volume = {48},
	copyright = {http://link.aps.org/licenses/aps-default-license},
	issn = {0163-1829, 1095-3795},
	url = {https://link.aps.org/doi/10.1103/PhysRevB.48.10345},
	doi = {10.1103/PhysRevB.48.10345},
	number = {14},
	urldate = {2026-03-29},
	journal = {Physical Review B},
	author = {White, Steven R.},
	month = oct,
	year = {1993},
	pages = {10345--10356},
}

@article{white_density_1992,
	volume = {69},
	copyright = {http://link.aps.org/licenses/aps-default-license},
	issn = {0031-9007},
	url = {https://link.aps.org/doi/10.1103/PhysRevLett.69.2863},
	doi = {10.1103/PhysRevLett.69.2863},
	number = {19},
	urldate = {2026-03-29},
	journal = {Physical Review Letters},
	author = {White, Steven R.},
	month = nov,
	year = {1992},
	pages = {2863--2866},
}

@article{Taffet2019,
	volume = {123},
	url = {https://doi.org/10.1021/acs.jpcb.9b04027},
	doi = {10.1021/acs.jpcb.9b04027},
	pages = {8628--8643},
	number = {41},
	journal = {The Journal of Physical Chemistry B},
	author = {Taffet, Elliot J. and Lee, Benjamin G. and Toa, Zi S. D. and Pace, Natalie and Rumbles, Garry and Southall, June and Cogdell, Richard J. and Scholes, Gregory D.},
	year = {2019},
}

@article{simpson_resonance_1955,
	volume = {77},
	issn = {0002-7863},
	url = {https://doi.org/10.1021/ja01628a020},
	doi = {10.1021/ja01628a020},
	number = {23},
	journal = {Journal of the American Chemical Society},
	publisher = {American Chemical Society},
	author = {Simpson, William T.},
	month = dec,
	year = {1955},
	pages = {6164--6168},
}

@article{Berkelbach2013,
	volume = {138},
	issn = {0021-9606, 1089-7690},
	url = {https://pubs.aip.org/jcp/article/138/11/114103/192428/Microscopic-theory-of-singlet-exciton-fission-II},
	doi = {10.1063/1.4794427},
	pages = {114103},
	number = {11},
	journal = {The Journal of Chemical Physics},
	author = {Berkelbach, Timothy C. and Hybertsen, Mark S. and Reichman, David R.},
	urldate = {2024-11-20},
	year = {2013},
	langid = {english},
}

@article{Santra2022,
	volume = {13},
	rights = {https://doi.org/10.15223/policy-029},
	issn = {1948-7185, 1948-7185},
	url = {https://pubs.acs.org/doi/10.1021/acs.jpclett.2c02000},
	doi = {10.1021/acs.jpclett.2c02000},
	pages = {6800--6805},
	number = {29},
	journal = {The Journal of Physical Chemistry Letters},
	shortjournal = {J. Phys. Chem. Lett.},
	author = {Santra, Supriyo and Ray, Jyotirmoy and Ghosh, Debashree},
	urldate = {2024-11-20},
	year = {2022},
	langid = {english},
}

@article{Young2020,
	volume = {53},
	rights = {https://doi.org/10.15223/policy-029},
	issn = {0001-4842, 1520-4898},
	url = {https://pubs.acs.org/doi/10.1021/acs.accounts.0c00397},
	doi = {10.1021/acs.accounts.0c00397},
	shorttitle = {Mixed Electronic States in Molecular Dimers},
	pages = {1957--1968},
	number = {9},
	journal = {Accounts of Chemical Research},
	shortjournal = {Acc. Chem. Res.},
	author = {Young, Ryan M. and Wasielewski, Michael R.},
	urldate = {2024-11-20},
	year = {2020},
	langid = {english},
}

@article{Barford2022,
	volume = {106},
	issn = {2469-9950, 2469-9969},
	url = {https://link.aps.org/doi/10.1103/PhysRevB.106.035201},
	doi = {10.1103/PhysRevB.106.035201},
	pages = {035201},
	number = {3},
	journal = {Physical Review B},
	shortjournal = {Phys. Rev. B},
	author = {Barford, William},
	urldate = {2024-11-20},
	year = {2022},
	langid = {english},
}

@article{Chambers2023,
	volume = {159},
	issn = {0021-9606, 1089-7690},
	url = {https://pubs.aip.org/jcp/article/159/8/084116/2908466/Theory-of-singlet-fission-in-carotenoid-dimers},
	doi = {10.1063/5.0155476},
	pages = {084116},
	number = {8},
	journal = {The Journal of Chemical Physics},
	author = {Barford, William and Chambers, Cameron A.},
	urldate = {2024-11-20},
	year = {2023},
	langid = {english},
}

@article{Chandross1999,
	volume = {59},
	rights = {http://link.aps.org/licenses/aps-default-license},
	issn = {0163-1829, 1095-3795},
	url = {https://link.aps.org/doi/10.1103/PhysRevB.59.4822},
	doi = {10.1103/PhysRevB.59.4822},
	pages = {4822--4838},
	number = {7},
	journal = {Physical Review B},
	shortjournal = {Phys. Rev. B},
	author = {Chandross, M. and Shimoi, Y. and Mazumdar, S.},
	urldate = {2024-11-20},
	year = {1999},
	langid = {english},
}

@article{Prodhan2017,
	volume = {96},
	rights = {https://link.aps.org/licenses/aps-default-license},
	issn = {2469-9950, 2469-9969},
	url = {https://link.aps.org/doi/10.1103/PhysRevB.96.075142},
	doi = {10.1103/PhysRevB.96.075142},
	pages = {075142},
	number = {7},
	journal = {Physical Review B},
	shortjournal = {Phys. Rev. B},
	author = {Prodhan, Suryoday and Ramasesha, S.},
	urldate = {2024-11-20},
	year = {2017},
	langid = {english},
}

@article{Aryanpour2015,
	volume = {119},
	issn = {1932-7447, 1932-7455},
	url = {https://pubs.acs.org/doi/10.1021/jp5124019},
	doi = {10.1021/jp5124019},
	pages = {6966--6979},
	number = {13},
	journal = {The Journal of Physical Chemistry C},
	shortjournal = {J. Phys. Chem. C},
	author = {Aryanpour, Karan and Shukla, Alok and Mazumdar, Sumit},
	urldate = {2024-11-20},
	year = {2015},
	langid = {english},
}

@article{Peng2024,
	volume = {160},
	issn = {0021-9606, 1089-7690},
	url = {https://pubs.aip.org/jcp/article/160/19/194304/3293997/Investigation-of-ultrafast-intermediate-states},
	doi = {10.1063/5.0200802},
	pages = {194304},
	number = {19},
	journal = {The Journal of Chemical Physics},
	author = {Peng, Bo and Wang, Ziyu and Jiang, Jiaming and Huang, Yifan and Liu, Weimin},
	urldate = {2024-11-27},
	year = {2024},
	langid = {english},
}

@article{Valentine2020,
	volume = {102},
	issn = {2469-9950, 2469-9969},
	url = {https://link.aps.org/doi/10.1103/PhysRevB.102.125107},
	doi = {10.1103/PhysRevB.102.125107},
	pages = {125107},
	number = {12},
	journal = {Physical Review B},
	shortjournal = {Phys. Rev. B},
	author = {Valentine, D. J. and Manawadu, D. and Barford, W.},
	urldate = {2024-11-28},
	year = {2020},
	langid = {english},
}

@article{Barford2023,
	volume = {14},
	rights = {https://creativecommons.org/licenses/by/4.0/},
	issn = {1948-7185, 1948-7185},
	url = {https://pubs.acs.org/doi/10.1021/acs.jpclett.3c02435},
	doi = {10.1021/acs.jpclett.3c02435},
	pages = {9842--9847},
	number = {44},
	journal = {The Journal of Physical Chemistry Letters},
	shortjournal = {J. Phys. Chem. Lett.},
	author = {Barford, William},
	urldate = {2025-01-22},
	year = {2023},
	langid = {english},
}

@article{Schollwock2005,
	volume = {77},
	number = {1},
    pages = {259},
    url = {https://doi.org/10.1103/RevModPhys.77.259},
	journal = {Rev. Mod. Phys.},
	author = {Schollwöck, U},
	year = {2005},
	langid = {english},
}

@article{Musser2015,
	volume = {137},
	rights = {http://pubs.acs.org/page/policy/authorchoice\_ccby\_termsofuse.html},
	issn = {0002-7863, 1520-5126},
	url = {https://pubs.acs.org/doi/10.1021/jacs.5b01130},
	doi = {10.1021/jacs.5b01130},
	pages = {5130--5139},
	number = {15},
	journal = {Journal of the American Chemical Society},
	shortjournal = {J. Am. Chem. Soc.},
	author = {Musser, Andrew J. and Maiuri, Margherita and Brida, Daniele and Cerullo, Giulio and Friend, Richard H. and Clark, Jenny},
	urldate = {2025-08-14},
	year = {2015},
	langid = {english},
}

@article{Ichert2025,
	volume = {27},
	issn = {1463-9076, 1463-9084},
	url = {https://xlink.rsc.org/?DOI=D4CP04445B},
	doi = {10.1039/D4CP04445B},
	pages = {2371--2380},
	number = {5},
	journal = {Physical Chemistry Chemical Physics},
	shortjournal = {Phys. Chem. Chem. Phys.},
	author = {Ichert, Alexandru G. and Barford, William},
	urldate = {2025-08-14},
	year = {2025},
	langid = {english},
}

@article{Manawadu2023,
	volume = {127},
	rights = {https://creativecommons.org/licenses/by/4.0/},
	issn = {1089-5639, 1520-5215},
	url = {https://pubs.acs.org/doi/10.1021/acs.jpca.2c07781},
	doi = {10.1021/acs.jpca.2c07781},
	pages = {1342--1352},
	number = {5},
	journal = {The Journal of Physical Chemistry A},
	shortjournal = {J. Phys. Chem. A},
	author = {Manawadu, Dilhan and Georges, Timothy N. and Barford, William},
	urldate = {2025-08-14},
	year = {2023},
	langid = {english},
}

@article{Santra2025,
	volume = {129},
	rights = {https://doi.org/10.15223/policy-029},
	issn = {1089-5639, 1520-5215},
	url = {https://pubs.acs.org/doi/10.1021/acs.jpca.5c00060},
	doi = {10.1021/acs.jpca.5c00060},
	shorttitle = {Unraveling the Geometrical Effects on Singlet Fission of Carotenoids},
	pages = {2738--2744},
	number = {11},
	journal = {The Journal of Physical Chemistry A},
	shortjournal = {J. Phys. Chem. A},
	author = {Santra, Supriyo and Ghosh, Debashree},
	urldate = {2025-08-14},
	year = {2025},
	langid = {english},
}

@article{Smith2010,
	volume = {110},
	issn = {0009-2665, 1520-6890},
	url = {https://pubs.acs.org/doi/10.1021/cr1002613},
	doi = {10.1021/cr1002613},
	pages = {6891--6936},
	number = {11},
	journal = {Chemical Reviews},
	shortjournal = {Chem. Rev.},
	author = {Smith, Millicent B. and Michl, Josef},
	urldate = {2025-08-14},
	year = {2010},
	langid = {english},
}

@article{Singh1965,
	volume = {42},
	issn = {0021-9606, 1089-7690},
	url = {https://pubs.aip.org/jcp/article/42/1/330/81704/Laser-Generation-of-Excitons-and-Fluorescence-in},
	doi = {10.1063/1.1695695},
	pages = {330--342},
	number = {1},
	journal = {The Journal of Chemical Physics},
	author = {Singh, S. and Jones, W. J. and Siebrand, W. and Stoicheff, B. P. and Schneider, W. G.},
	urldate = {2025-08-14},
	year = {1965},
	langid = {english},
}

@article{Shockley1961,
	volume = {32},
	issn = {0021-8979, 1089-7550},
	url = {https://pubs.aip.org/jap/article/32/3/510/505950/Detailed-Balance-Limit-of-Efficiency-of-p-n},
	doi = {10.1063/1.1736034},
	pages = {510--519},
	number = {3},
	journal = {Journal of Applied Physics},
	author = {Shockley, William and Queisser, Hans J.},
	urldate = {2025-08-14},
	year = {1961},
	langid = {english},
}

@article{Chagas2025,
	volume = {27},
	issn = {1463-9076, 1463-9084},
	url = {https://xlink.rsc.org/?DOI=D5CP00339C},
	doi = {10.1039/D5CP00339C},
	shorttitle = {Low-lying excited states of linear all- \textit{trans} polyenes},
	pages = {7916--7928},
	number = {15},
	journal = {Physical Chemistry Chemical Physics},
	shortjournal = {Phys. Chem. Chem. Phys.},
	author = {Chagas, Julio C. V. and F. Dos Santos, Luan G. and Nieman, Reed and Aquino, Adelia J. A. and Do Monte, Silmar A. and Plasser, Felix and Szalay, Péter G. and Lischka, Hans and Machado, Francisco B. C.},
	urldate = {2025-09-12},
	year = {2025},
	langid = {english},
}

@article{Schulten1976,
	volume = {64},
	issn = {0021-9606, 1089-7690},
	url = {https://pubs.aip.org/jcp/article/64/11/4422/443544/Correlation-effects-in-the-spectra-of-polyenes},
	doi = {10.1063/1.432121},
	pages = {4422--4441},
	number = {11},
	journal = {The Journal of Chemical Physics},
	author = {Schulten, K. and Ohmine, I. and Karplus, M.},
	urldate = {2026-01-14},
	year = {1976},
	langid = {english},
}

@book{WBBook,
	edition = {2nd},
	title = {Electronic and Optical Properties of Conjugated Polymers},
	publisher = {Oxford University Press},
	author = {Barford, William},
    year = {2013}
}

@article{Starcke2006,
	volume = {329},
	rights = {https://www.elsevier.com/tdm/userlicense/1.0/},
	issn = {03010104},
	url = {https://linkinghub.elsevier.com/retrieve/pii/S0301010406004022},
	doi = {10.1016/j.chemphys.2006.07.020},
	pages = {39--49},
	number = {1},
	journal = {Chemical Physics},
	shortjournal = {Chemical Physics},
	author = {Starcke, Jan Hendrik and Wormit, Michael and Schirmer, Jochen and Dreuw, Andreas},
	urldate = {2026-02-10},
	year = {2006},
	langid = {english},
}

@article{Boguslawski2019,
	volume = {15},
	issn = {1549-9618, 1549-9626},
	url = {https://pubs.acs.org/doi/10.1021/acs.jctc.8b01053},
	doi = {10.1021/acs.jctc.8b01053},
	pages = {18--24},
	number = {1},
	journal = {Journal of Chemical Theory and Computation},
	shortjournal = {J. Chem. Theory Comput.},
	author = {Boguslawski, Katharina},
	urldate = {2026-02-24},
	year = {2019},
	langid = {english},
}

@article{Magne2025,
	volume = {15},
	issn = {2045-2322},
	url = {https://www.nature.com/articles/s41598-025-88220-z},
	doi = {10.1038/s41598-025-88220-z},
	pages = {5593},
	number = {1},
	journal = {Scientific Reports},
	shortjournal = {Sci Rep},
	author = {Magne, Chloe and Veremeienko, Vasyl and Bercy, Roxanne and Ha-Thi, Minh-Huong and Arteni, Ana A. and Pascal, Andrew A. and Vengris, Mikas and Pino, Thomas and Robert, Bruno and Llansola-Portoles, Manuel J.},
	urldate = {2026-03-12},
	year = {2025},
	langid = {english},
}

@article{Kundu2021,
	volume = {12},
	rights = {https://doi.org/10.15223/policy-029},
	issn = {1948-7185, 1948-7185},
	url = {https://pubs.acs.org/doi/10.1021/acs.jpclett.0c03301},
	doi = {10.1021/acs.jpclett.0c03301},
	pages = {1468--1474},
	number = {5},
	journal = {The Journal of Physical Chemistry Letters},
	shortjournal = {J. Phys. Chem. Lett.},
	author = {Kundu, Arup and Dasgupta, Jyotishman},
	urldate = {2026-03-19},
	year = {2021},
	langid = {english},
}

@article{Mulliken1949,
	volume = {17},
	issn = {0021-9606, 1089-7690},
	url = {https://pubs.aip.org/jcp/article/17/12/1248/72671/Formulas-and-Numerical-Tables-for-Overlap},
	doi = {10.1063/1.1747150},
	pages = {1248--1267},
	number = {12},
	journal = {The Journal of Chemical Physics},
	author = {Mulliken, R. S. and Rieke, C. A. and Orloff, D. and Orloff, H.},
	urldate = {2026-03-23},
	year = {1949},
	langid = {english},
}

@article{Mulliken1955_2,
	volume = {23},
	issn = {0021-9606, 1089-7690},
	url = {https://pubs.aip.org/jcp/article/23/10/1841/76998/Electronic-Population-Analysis-on-LCAO-MO},
	doi = {10.1063/1.1740589},
	pages = {1841--1846},
	number = {10},
	journal = {The Journal of Chemical Physics},
	author = {Mulliken, R. S.},
	urldate = {2026-03-25},
	year = {1955},
	langid = {english},
}

@article{Santra2026,
	volume = {22},
	rights = {https://doi.org/10.15223/policy-029},
	issn = {1549-9618, 1549-9626},
	url = {https://pubs.acs.org/doi/10.1021/acs.jctc.6c00429},
	doi = {10.1021/acs.jctc.6c00429},
	pages = {4666--4674},
	number = {9},
	journal = {Journal of Chemical Theory and Computation},
	shortjournal = {J. Chem. Theory Comput.},
	author = {Santra, Supriyo and Bose, Amartya and Ghosh, Debashree},
	urldate = {2026-05-26},
	year = {2026},
	langid = {english},
}

@article{Wang2026,
	issn = {1755-4330, 1755-4349},
	url = {https://www.nature.com/articles/s41557-026-02186-7},
	journal = {Nature Chemistry},
    volume = {18},
    pages = {1611-1619},
	shortjournal = {Nat. Chem.},
	author = {Wang, Shuangqing and Sutherland, George A. and Pidgeon, James P. and Šmitran, Mateja and Swainsbury, David J. K. and Martin, Elizabeth C. and Vasilev, Cvetelin and Hitchcock, Andrew and Gillard, Daniel J. and Kumar Venkatraman, Ravi and Chekulaev, Dimitri and Tartakovskii, Alexander I. and Hunter, C. Neil and Clark, Jenny},
	urldate = {2026-07-14},
	year = {2026},
	langid = {english}
}

@article{Wang2010,
	volume = {132},
	issn = {0002-7863, 1520-5126},
	url = {https://pubs.acs.org/doi/10.1021/ja102851m},
	doi = {10.1021/ja102851m},
	pages = {13988--13991},
	number = {40},
	journal = {Journal of the American Chemical Society},
	shortjournal = {J. Am. Chem. Soc.},
	author = {Wang, Chen and Tauber, Michael J.},
	urldate = {2026-09-28},
	year = {2010},
	langid = {english},
}

@article{Smith2013,
	volume = {64},
	issn = {0066-426X, 1545-1593},
	url = {https://www.annualreviews.org/doi/10.1146/annurev-physchem-040412-110130},
	doi = {10.1146/annurev-physchem-040412-110130},
	pages = {361--386},
	number = {1},
	journal = {Annual Review of Physical Chemistry},
	shortjournal = {Annu. Rev. Phys. Chem.},
	author = {Smith, Millicent B. and Michl, Josef},
	urldate = {2026-09-30},
	year = {2013},
	langid = {english},
}

@article{Hannah2006,
	volume = {100},
	issn = {0021-8979, 1089-7550},
	url = {https://pubs.aip.org/jap/article/100/7/074510/342873/Solar-conversion-efficiency-of-photovoltaic-and},
	doi = {10.1063/1.2356795},
	pages = {074510},
	number = {7},
	journal = {Journal of Applied Physics},
	author = {Hanna, M. C. and Nozik, A. J.},
	urldate = {2026-09-30},
	year = {2006},
	langid = {english},
}

@article{Monahan2015,
	volume = {66},
	issn = {0066-426X, 1545-1593},
	url = {https://www.annualreviews.org/doi/10.1146/annurev-physchem-040214-121235},
	doi = {10.1146/annurev-physchem-040214-121235},
	pages = {601--618},
	number = {1},
	journal = {Annual Review of Physical Chemistry},
	shortjournal = {Annu. Rev. Phys. Chem.},
	author = {Monahan, N. and Zhu, X.-Y.},
	urldate = {2026-09-30},
	year = {2015},
	langid = {english},
}

@article{Schulten1972,
	volume = {14},
	issn = {00092614},
	url = {https://linkinghub.elsevier.com/retrieve/pii/0009261472801209},
	doi = {10.1016/0009-2614(72)80120-9},
	pages = {305--309},
	number = {3},
	journal = {Chemical Physics Letters},
	shortjournal = {Chemical Physics Letters},
	author = {Schulten, K. and Karplus, M.},
	urldate = {2026-09-30},
	year = {1972},
	langid = {english},
}

@article{Hayden1986,
	volume = {34},
	rights = {http://link.aps.org/licenses/aps-default-license},
	issn = {0163-1829},
	url = {https://link.aps.org/doi/10.1103/PhysRevB.34.5484},
	doi = {10.1103/PhysRevB.34.5484},
	pages = {5484--5497},
	number = {8},
	journal = {Physical Review B},
	shortjournal = {Phys. Rev. B},
	author = {Hayden, G. W. and Mele, E. J.},
	urldate = {2026-09-30},
	year = {1986},
	langid = {english},
}

@article{Ichert2026,
	volume = {28},
	issn = {1463-9076, 1463-9084},
	url = {http://pubs.rsc.org/CP/article-lookup/doi/10.1039/D6CP01090C},
	doi = {10.1039/D6CP01090C},
	pages = {18116--18123},
	number = {29},
	journal = {Physical Chemistry Chemical Physics},
	shortjournal = {Phys. Chem. Chem. Phys.},
	author = {Ichert, Alexandru G. and Barford, William},
	urldate = {2026-09-30},
	year = {2026},
}

@article{Quaranta2021,
	volume = {23},
	issn = {1463-9076, 1463-9084},
	url = {https://pubs.rsc.org/cp/article/23/8/4768-4776/680711},
	doi = {10.1039/D0CP04493H},
	pages = {4768--4776},
	number = {8},
	journal = {Physical Chemistry Chemical Physics},
	shortjournal = {Phys. Chem. Chem. Phys.},
	author = {Quaranta, Annamaria and Krieger-Liszkay, Anja and Pascal, Andrew A. and Perreau, François and Robert, Bruno and Vengris, Mikas and Llansola-Portoles, Manuel J.},
	urldate = {2026-09-30},
	year = {2021},
}

@article{Georges2026,
	volume = {130},
	rights = {https://creativecommons.org/licenses/by/4.0/},
	issn = {1089-5639, 1520-5215},
	url = {https://pubs.acs.org/doi/10.1021/acs.jpca.6c00402},
	doi = {10.1021/acs.jpca.6c00402},
	pages = {2765--2777},
	number = {13},
	journal = {The Journal of Physical Chemistry A},
	shortjournal = {J. Phys. Chem. A},
	author = {Georges, Timothy N. and Summerley, Louis and Runeson, Johan E. and Barford, William},
	urldate = {2026-09-30},
	year = {2026},
	langid = {english},

}

@article{Manawadu2022,
	volume = {13},
	rights = {https://creativecommons.org/licenses/by/4.0/},
	issn = {1948-7185, 1948-7185},
	url = {https://pubs.acs.org/doi/10.1021/acs.jpclett.1c03812},
	doi = {10.1021/acs.jpclett.1c03812},
	pages = {1344--1349},
	number = {5},
	journal = {The Journal of Physical Chemistry Letters},
	shortjournal = {J. Phys. Chem. Lett.},
	author = {Manawadu, Dilhan and Valentine, Darren J. and Marcus, Max and Barford, William},
	urldate = {2026-09-30},
	year = {2022},
}

@article{Portoles2018,
	volume = {20},
	issn = {1463-9076, 1463-9084},
	url = {https://pubs.rsc.org/cp/article/20/13/8640-8646/561856},
	doi = {10.1039/C7CP08460A},
	pages = {8640--8646},
	number = {13},
	journal = {Physical Chemistry Chemical Physics},
	shortjournal = {Phys. Chem. Chem. Phys.},
	author = {Llansola-Portoles, M. J. and Redeckas, K. and Streckaité, S. and Ilioaia, C. and Pascal, A. A. and Telfer, A. and Vengris, M. and Valkunas, L. and Robert, B.},
	urldate = {2026-09-30},
	year = {2018},
	langid = {english},
}

@article{Swenberg1968,
	volume = {2},
	issn = {0009-2614},
	url = {https://www.sciencedirect.com/science/article/pii/0009261468800879},
	doi = {https://doi.org/10.1016/0009-2614(68)80087-9},
	pages = {327--328},
	number = {5},
	journal = {Chemical Physics Letters},
	author = {Swenberg, C. E. and Stacy, W. T.},
	year = {1968},
}

@article{Schollwock2011,
    journal = {Annals of Physics},
    volume = {326},
    number = {1},
    pages = {96-192},
    year = {2011},
    issn = {0003-4916},
    doi = {https://doi.org/10.1016/j.aop.2010.09.012},
    url = {https://www.sciencedirect.com/science/article/pii/S0003491610001752},
    author = {Ulrich Schollwöck},
}

\end{document}



\title{Supporting Information\\Geometry-Dependent Electronic Coupling and\\Singlet Fission Pathways in Carotenoid Dimers}

\author{Alexandru G. Ichert}
\affiliation{Department of Chemistry, Physical and Theoretical Chemistry Laboratory, University of Oxford, Oxford, OX1 3QZ, United Kingdom}
\affiliation{Linacre College, University of Oxford, Oxford, OX1 3JA, United Kingdom}
\email{alexandru.ichert@chem.ox.ac.uk}

\author{George A. Smith}
\affiliation{Department of Chemistry, Physical and Theoretical Chemistry Laboratory, University of Oxford, Oxford, OX1 3QZ, United Kingdom}
\affiliation{Worcester College, University of Oxford, Oxford, OX1 2BH, United Kingdom}

\author{William Barford}
\affiliation{Department of Chemistry, Physical and Theoretical Chemistry Laboratory, University of Oxford, Oxford, OX1 3QZ, United Kingdom}
\email{william.barford@chem.ox.ac.uk}


\maketitle
\section{DMRG and MPS wavefunction mapping}\label{app:DMRG_MPS}

We use the infinite chain DMRG algorithm to grow the size of the system to the desired size, after which sweeping is used to ensure convergence of the results. See Refs.~\cite{Schollwock2005,Schollwock2011} for more details of DMRG and Section~\ref{conv_test} for more details of the convergence tests. 

\begin{figure}[b]
    \centering
    \includegraphics[width=0.75\linewidth]{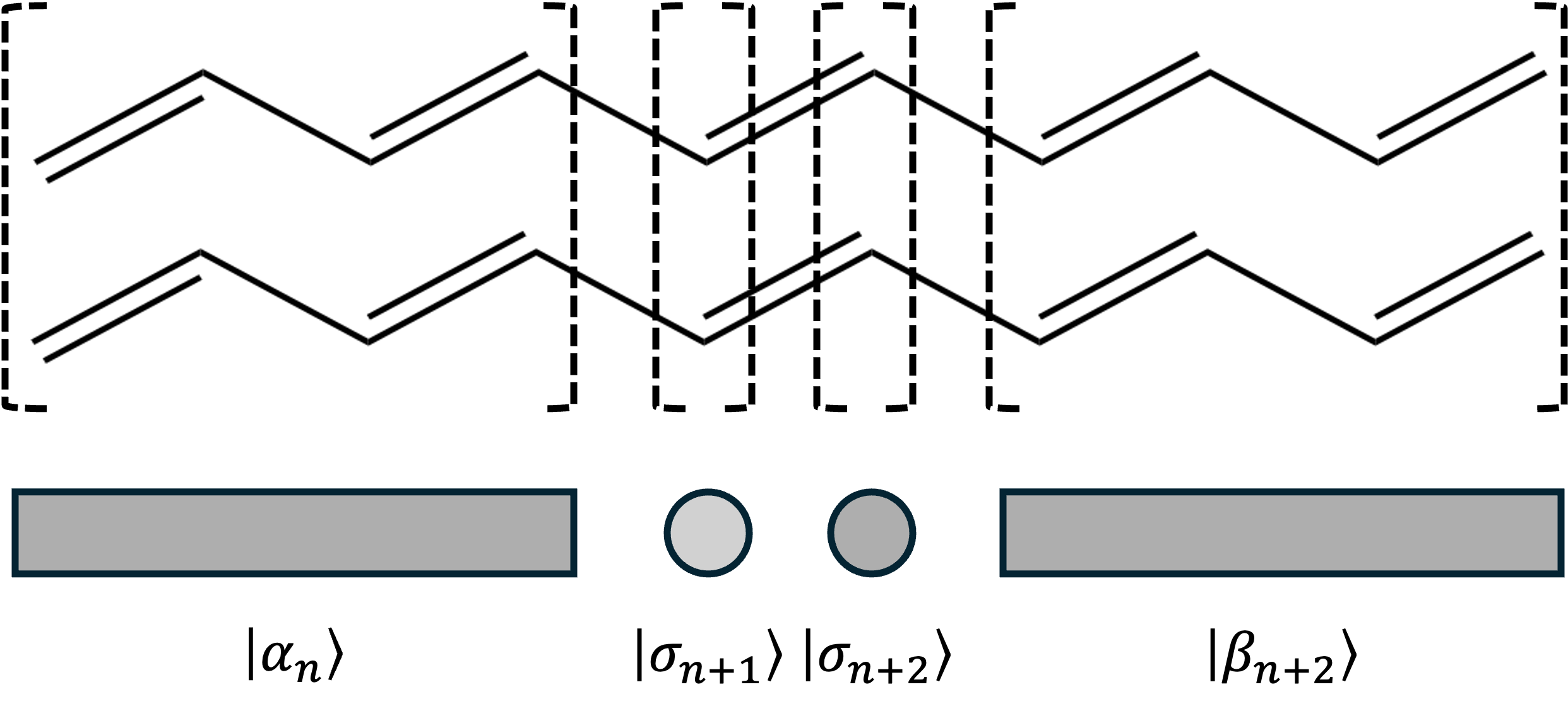}
    \caption{The block structure of the DMRG wavefunction. The primitive blocks are formed by augmenting  two individual sites on separate chains. The system can then be thought of as a quasi-one-dimensional system where the `site' is replaced by a  primitive block of two sites.}
    \label{fig:DMRG_blocks}
\end{figure}

A system of two chains is a quasi-one-dimensional system, where the primitive block is composed of two sites, one on each chain (as shown in Fig.~\ref{fig:DMRG_blocks}). We can therefore maintain the usual four-block structure of the wavefunction (i.e., two DMRG blocks and two primitive blocks). We write the wavefunction for two chains of $N$ sites each as
\begin{equation}
    |\Psi_{N}\rangle = \sum_{\alpha_n\sigma_{n+1}\sigma_{n+2}\beta_{n+2}} \Psi_{\alpha_n\sigma_{n+1}\sigma_{n+2}\beta_{n+2}}|\alpha_n\rangle |\sigma_{n+1}\rangle |\sigma_{n+2}\rangle |\beta_{n+2}\rangle,
\end{equation}
where the kets $|\sigma_m\rangle$ describe the $m$th sites on both chains simultaneously, i.e., $|\sigma_m\rangle \equiv |\sigma^{(1)}_m\rangle |\sigma^{(2)}_m\rangle$. For four possible site states, $\left\{ |0\rangle,|\uparrow\rangle,|\downarrow\rangle,|\uparrow\downarrow\rangle  \right\}$, there are 16 total primitive block states. The DMRG blocks are described by $|\alpha_n\rangle$ and $|\beta_{n+2}\rangle$, where
\begin{widetext}
\begin{equation}
    |\alpha_n\rangle = \sum_{\alpha_1,\alpha_2,\ldots,\alpha_{n-1}} \sum_{\sigma_1, \sigma_2,\ldots,\sigma_n}
    A_{\alpha_1}^{\sigma_1} A_{\alpha_1,\alpha_2}^{\sigma_2}\ldots A_{\alpha_{n-1},\alpha_{n}}^{\sigma_n} |\sigma_1 \sigma_2\ldots \sigma_n\rangle
\end{equation}
\end{widetext}
and 
\begin{widetext}
\begin{equation}
    |\beta_{n+2}\rangle = \sum_{\beta_{n+3},\ldots,\beta_{N-2},\beta_{N-1}} \sum_{\sigma_{n+3},\ldots,\sigma_{N-1}, \sigma_{N}}
    B_{\beta_{n+2},\beta_{n+3}}^{\sigma_{n+3}}\ldots B_{\beta_{N-2},\beta_{N-1}}^{\sigma_{N-1}}B_{\beta_{N-1}}^{\sigma_{N}} |\sigma_{n+3} \ldots \sigma_{N-1}\sigma_{N}\rangle.
\end{equation}
\end{widetext}
$A_{\alpha_{m-1},\alpha_m}^{\sigma_m}$ is the matrix product state (MPS) tensor corresponding to the $m$th carbon atom in the left block and $B_{\beta_{m-1},\beta_{m}}^{\sigma_m}$ is the MPS tensor corresponding to the $m$th carbon atom in the right block, as seen in Fig.~\ref{fig:DMRG_blocks}.

Once both the coupled and uncoupled eigenstates are calculated, we find the projections of the coupled chain states onto  the uncoupled chain states. This procedure is straightforward, as once written in terms of their respective MPS tensors, all wavefunctions are expressed in the same (site) basis. A similar procedure was employed in Ref.~\cite{Ichert2026} to determine the triplet-pair character of the $2^1\textrm{A}_\textrm{g}$ state of polyenes.

\section{DMRG Convergence Tests}\label{conv_test}
Table~\ref{si:conv} shows the convergence of energy with the number of block states (i.e., the bond dimension) and sweeps for two coupled lycopene chains in the vertically stacked conformation at 3.5 \AA~chain separation. The results presented in the paper were calculated using 1,000 block states, with the exception of the $\textrm{T}_1\textrm{T}_2$ uncoupled triplet-pairs, which were calculated using 850 block states. These choices were made to avoid prohibitive computational costs for the calculations presented in the main paper.

\begin{table}[h!]
\centering
\begin{tabular}{|c|c|c|c|c|}
\hline
Bond Dimension & No. Sweeps & Ex. Energy 1 & Ex. Energy 2 & Ex. Energy 3 \\
    \hline
694  & 0 & 2.4765 & 2.4956 & 2.6483 \\
     & 1 & 2.4700 & 2.4910 & 2.6460 \\
     & 2 & 2.4696 & 2.4907 & 2.6457 \\
    \hline
893  & 0 & 2.4600 & 2.4810 & 2.6381 \\
     & 1 & 2.4561 & 2.4786 & 2.6359 \\
     & 2 & 2.4559 & 2.4784 & 2.6357 \\
     \hline
1072 & 0 & 2.4527 & 2.4742 & 2.6337 \\
     & 1 & 2.4493 & 2.4719 & 2.6316 \\
     & 2 & 2.4492 & 2.4718 & 2.6315 \\
     \hline
1227 & 0 & 2.4484 & 2.4702 & 2.6310 \\
     & 1 & 2.4465 & 2.4689 & 2.6294 \\
     & 2 & 2.4463 & 2.4688 & 2.6293 \\
     \hline
1446 & 0 & 2.4484 & 2.4702 & 2.6310 \\
     & 1 & 2.4465 & 2.4689 & 2.6294 \\
     & 2 & 2.4463 & 2.4688 & 2.6293 \\
     \hline
1667 & 0 & 2.4443 & 2.4662 & 2.6280 \\
     & 1 & 2.4433 & 2.4656 & 2.6267 \\
     & 2 & 2.4432 & 2.4655 & 2.6267 \\
     \hline
1844 & 0 & 2.4433 & 2.4653 & 2.6273 \\
     & 1 & 2.4426 & 2.4649 & 2.6261 \\
     & 2 & 2.4426 & 2.4648 & 2.6261 \\
     \hline
\end{tabular}
\caption{Convergence tests for the energies of the three lowest excited states of vertically stacked lycopene at 3.5 \AA~interchain distance.}\label{si:conv}
\end{table}

In all cases of bond dimension, the excitation energies converge to two decimal places very quickly with the number of sweeps, so the results after two sweeps were used in the main paper. Likewise, our choice of bond dimension of 1,000 gives us excitation energies which are within 0.01 eV of extrapolated values.

\section{Molecular-Exciton, Valence-Bond Model of Electronic States in Polyenes}

The molecular-exciton, valence-bond approach, introduced by Simpson \cite{simpson_resonance_1955}, treats excitations in a conjugated polymer as being built from excitations within the monomers. The polymer linkage is then treated as a perturbation onto the set of independent simple monomer excitations.
In this framework, the polyene excited states are expressed as direct products of ethylene monomer states. The ethylene eigenstates and their energies are derived exactly from the PPP Hamiltonian defined in Section II of the main paper. This approach was used  by Chandross \textit{et al.}\ \cite{Chandross1999} and Barford\ \cite{Barford2022} to characterize the excited states of polyenes.

We first define the primitive basis states (Section S3.A), then -- after defining the effect of symmetry operations on these basis states (Section S3.B) --  we define the symmetry adapted basis states (Section S3.C). The symmetry-allowed nonzero matrix elements are listed in Section S3.D, from which the selection rules described in Section V.B of the main paper follow.


\subsection{Primitive basis states}\label{RM.1}

To construct this minimal model, we restrict our choice of electronic configurations to those that have been shown to be the dominant contributors to the full CI wavefunctions \cite{Chandross1999}. For the ground state, this is simply,
\begin{equation}
    \ket{\textrm{GS}} = \bigotimes_i\ket{\textrm{GS}_i},
\end{equation}
where the product is over all ethylene units.  The simplest excitations are equivalent to one-monomer excited states, the lowest being the triplet,
\begin{equation}\label{Eq:16}
    \ket{\textrm{T},M_s;j}=\ket{\textrm{T}_j,M_s}\bigotimes_{i\neq j}\ket{\textrm{GS}_i},
\end{equation}
where $M_s$ is the spin magnetic quantum number (1, 0, or -1).

Next is the  singlet excitation,
\begin{equation}\label{Eq:17}
    \ket{\textrm{FE};j}=\ket{\textrm{FE}_j}\bigotimes_{i \neq j}\ket{\textrm{GS}_i},
\end{equation}
where we denote the singlet ethylene monomer excitation as a Frenkel exciton basis state as it consists of an electron-hole pair with minimal spatial separation.

To describe the ability of these excitons to delocalize across the system, we consider the intrachain charge-transfer exciton  with separated electron ($\ket{e^-;i}$) and hole ($\ket{h^+;j}$),
\begin{equation}\label{Eq:18}
    \ket{\textrm{EX}_{1};i,j} = \ket{e^-;i}\ket{h^+;j}\bigotimes_{k\neq i,j}\ket{\textrm{GS}_k}.
\end{equation}

The intrachain singlet triplet-pair excitation is defined as
\begin{equation}\label{Eq:19}
    \ket{\textrm{iTT};i,j}=\frac{1}{\sqrt{3}}(\ket{\textrm{T},1;i}\ket{\textrm{T},-1;j}-\ket{\textrm{T},0;i}\ket{\textrm{T},0;j}+\ket{\textrm{T},-1;i}\ket{\textrm{T},1;j}),
\end{equation}
where the triplet excitations act as hard-core bosons forbidden from occupying the same site.

\begin{figure}
    \centering
    \includegraphics[width=1\linewidth]{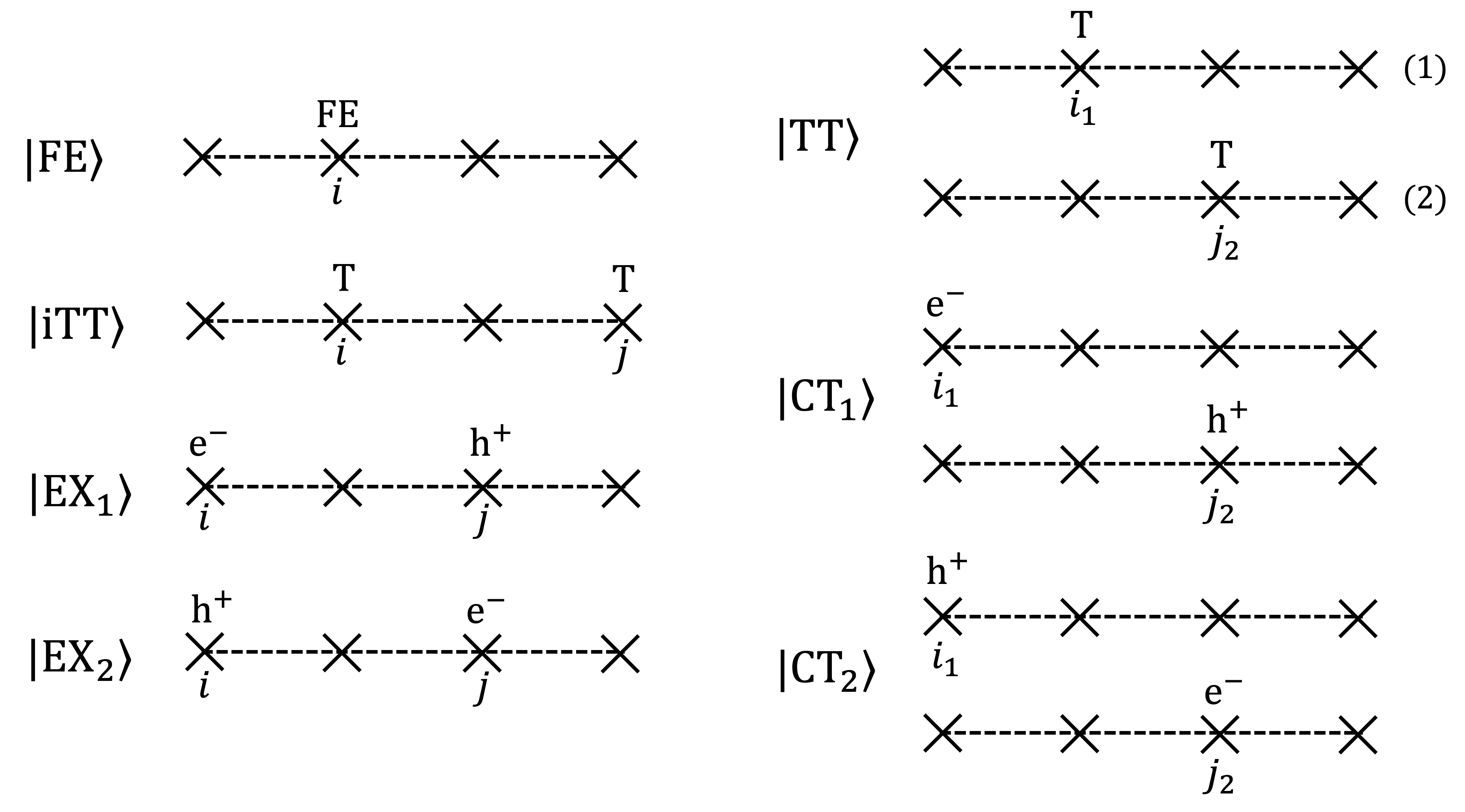}
    \caption{Schematic diagram of the basis states of the molecular-exciton, valence-bond model. The left-hand panel illustrates intrachain basis states, and the right-hand panel illustrates interchain basis states. A $\times$ symbol represents an ethylene dimer and a  dashed line represents a single bond.
     FE, EX and iTT denote the Frenkel exciton, intrachain charge-transfer exciton and intrachain triplet-pair, respectively; CT and TT denote the  interchain charge-transfer exciton and interchain triplet-pair, respectively.
     }
    \label{fig:RM_basis}
\end{figure}

There are two additional types of basis state that are required to describe the dimer structure. The first is the interchain singlet charge-transfer state between index \textit{i} on chain \textit{x} and \textit{j} on chain $\bar{x}$,
\begin{equation}\label{Eq:20}
    \ket{\textrm{CT}_1;i_x,j_{\bar{x}}}=\ket{e^-;i_x}\ket{h^+;j_{\bar{x}}} \bigotimes_{k,x\neq i(x),j(\bar{x})} \ket{\textrm{GS}_{k_x}}.
\end{equation}
These allow for hopping of excitons across chains, and the electron and hole of a CT basis state are allowed to hop along their respective chains (in analogy to EX).

The second new basis state is the interchain singlet triplet-pair in which each chain has a single triplet excitation,
\begin{equation}\label{Eq:21}
    \ket{\textrm{TT};i_x,j_{\bar{x}}}=\frac{1}{\sqrt{3}}\big(\ket{\textrm{T},1;i_x}\ket{\textrm{T},-1;j_{\bar{x}}}
    -\ket{\textrm{T},0;i_x}\ket{\textrm{T},0;j_{\bar{x}}}
    +\ket{\textrm{T},-1;i_x}\ket{\textrm{T},1;j_{\bar{x}}}\big),
\end{equation}
where $\ket{\textrm{T};i_x}$ is a triplet excitation on site \textit{i} of chain \textit{x}.

These  basis states are illustrated in Fig.\ \ref{fig:RM_basis} and their PPP model derived excitation energies are given in Section S3.E.

\subsection{Role of symmetry operations}

This section describes the role of the electron-hole and spatial symmetry operations on the primitive basis states.

\subsubsection{Electron-hole symmetry, $\hat{J}$}

\begin{equation}
\hat{J} | \textrm{FE}; i \rangle = + | \textrm{FE}; i \rangle,
\end{equation}

\begin{equation}
\hat{J} | \textrm{iTT}; i,j \rangle = - | \textrm{iTT}; i,j \rangle,
\end{equation}

\begin{equation}
\hat{J} | \textrm{TT}; i,j \rangle = - | \textrm{TT}; i,j \rangle,
\end{equation}
and
\begin{equation}
\hat{J} | \textrm{EX}_1^x; i,j \rangle = - | \textrm{EX}_2^x; i,j \rangle,
\end{equation}
where $x$ is the chain index.

\begin{equation}
\hat{J} | \textrm{CT}_1; i,j \rangle = + | \textrm{CT}_2; i,j \rangle; \textrm{ vertically stacked,}
\end{equation}
and
\begin{equation}
\hat{J} | \textrm{CT}_1; i,j \rangle = - | \textrm{CT}_2; i,j \rangle; \textrm{ slip-stacked.}
\end{equation}

Notice the opposite sign for the operation of $\hat{J}$ on $| \textrm{CT}_1; i,j \rangle$ for the vertically- and slip-staked geometries.

\subsubsection{Reflection symmetry, $\hat{\sigma}_{h}$, for vertically stacked dimers}

\begin{equation}
\hat{\sigma}_{h} | \textrm{EX}_y^x; i,j \rangle =  | \textrm{EX}_y^{\bar{x}}; i,j \rangle,
\end{equation}
where $\bar{x}$ is the opposite chain index to $x$ and $y =$ 1 or 2,
and
\begin{equation}
\hat{\sigma}_{h} | \textrm{CT}_1; i,j \rangle =  -| \textrm{CT}_2; j,i \rangle.
\end{equation}

\subsubsection{Inversion symmetry, $\hat{I}$, for slip-stacked dimers} 

\begin{equation}
\hat{I} | \textrm{EX}_y^x; i,j \rangle =  | \textrm{EX}_y^{\bar{x}}; N-j+1,N-i+1 \rangle,
\end{equation}
where $\bar{x}$ is the opposite chain index to $x$ and $y =$ 1 or 2,
and
\begin{equation}
 \hat{I} | \textrm{CT}_1; i,j \rangle =  | \textrm{CT}_2;N-j+1,N-i+1 \rangle.
\end{equation}

\subsection{Symmetry-adapted basis states}

\subsubsection{Electron-hole adapted (labeled by the $\pm$ superscript)}

\begin{equation}
 | \textrm{EX}^{x, \pm}; i,j \rangle = \frac{1}{\sqrt{2}}\left ( | \textrm{EX}_1^x; i,j \rangle \mp  | \textrm{EX}^x_2; i,j \rangle \right),
\end{equation}
where $x$ is the chain index.

\begin{equation}
 | \textrm{CT}^{\pm}; i,j \rangle = \frac{1}{\sqrt{2}}\left ( | \textrm{CT}_1; i,j \rangle \pm  | \textrm{CT}_2; i,j \rangle \right); \textrm{ vertically stacked,}
\end{equation}
and
\begin{equation}
 | \textrm{CT}^{\pm}; i,j \rangle = \frac{1}{\sqrt{2}}\left ( | \textrm{CT}_1; i,j \rangle \mp  | \textrm{CT}_2; i,j \rangle \right); \textrm{ slip-stacked.}
\end{equation}

\subsubsection{Electron-hole and spatially adapted  (labeled by the $\pm$ subscript)}

\begin{equation}\label{Eq:A11}
 | \textrm{FE}^{+}_{\pm}; i \rangle = \frac{1}{\sqrt{2}}\left ( | \textrm{FE}^{1,+}; i \rangle \pm  | \textrm{FE}^{2,+}; i\rangle \right)
\end{equation}
and
\begin{equation}\label{Eq:A12}
 | \textrm{EX}^{\pm}_{\pm}; i,j \rangle = \frac{1}{\sqrt{2}}\left ( | \textrm{EX}^{1,\pm}; i,j \rangle \pm  | \textrm{EX}^{2,\pm}; i,j \rangle \right).
\end{equation}

\vfill\pagebreak

Vertical- and slip-stacked:

\begin{equation}
 | \textrm{CT}^{\pm}_{\pm}; i,j \rangle = \frac{1}{\sqrt{2}}\left ( | \textrm{CT}^{\pm}; i,j \rangle \mp  | \textrm{CT}^{\pm}; j,i \rangle \right).
\end{equation}

In addition, for slip-stacked:

\begin{equation}
 | \textrm{CT}^{\pm}_{\pm}; i+1,j \rangle = \frac{1}{\sqrt{2}}\left ( | \textrm{CT}^{\pm}; i+1,j \rangle \mp  | \textrm{CT}^{\pm}; j,i-1 \rangle \right)
\end{equation}
and
\begin{equation}\label{Eq:A15}
 | \textrm{CT}^{\pm}_{\pm}; i,j-1 \rangle = \frac{1}{\sqrt{2}}\left ( | \textrm{CT}^{\pm}; i,j-1 \rangle \mp  | \textrm{CT}^{\pm}; j+1,i \rangle \right).
\end{equation}

\subsection{Couplings}\label{RM.couplings}

This section lists the nonzero matrix elements between the symmetry-adapted states, defined in Section S3.C. They are used to derive the interchain selection rules listed in Section V.B of the main paper. The parameters are defined in Section S3.E.

\subsubsection{Intrachain}

\begin{equation}
  \langle \textrm{FE}^+;i \pm 1 |\hat{H}| \textrm{FE}^+; i \rangle =-\beta^2 (V_1 - 2V_2 + V_3),
\end{equation}

\begin{equation}
  \langle \textrm{EX}^{\pm};i+1, j |\hat{H}| \textrm{EX}^{\pm}; i,j  \rangle = \frac{(\alpha +\beta)^2}{2}t_s,
\end{equation}

\begin{equation}
  \langle \textrm{iTT}^-;i \pm 1,j |\hat{H}| \textrm{iTT}^-; i,j \rangle =- \frac{\alpha^2t_s^2}{(E_{\textrm{EX}}^{\textrm{NN}}-E_{\textrm{T}})},
\end{equation}

\begin{equation}
  \langle \textrm{EX}^{+};i, i+1 |\hat{H}| \textrm{FE}^{+}; i  \rangle = \frac{(\alpha +\beta)}{2}t_s,
\end{equation}
and
\begin{equation}
  \langle \textrm{EX}^{-};i, i+1 |\hat{H}| \textrm{iTT}^-; i \pm 1  \rangle = -\sqrt{\frac{3}{4}} t_s.
\end{equation}

\subsubsection{Interchain: Vertically stacked}

\begin{equation}\label{Eq:A20}
  \langle \textrm{CT}^{+}_-;i,i+1 |\hat{H}| \textrm{FE}^+_- ;i  \rangle = -\sqrt{2}(\alpha + \beta) t_{\textrm{inter}},
\end{equation}

\begin{equation}\label{Eq:A21}
  \langle \textrm{CT}^{\pm}_-;i,j |\hat{H}| \textrm{EX}^{\pm}_- ;i,j  \rangle = -{2}(\alpha + \beta)^2 t_{\textrm{inter}},
\end{equation}
and
\begin{equation}\label{Eq:A211}
  \langle \textrm{iTT}_+^-;i,j |\hat{H}| \textrm{TT}_+^- ;i,j  \rangle = - \frac{2t_\textrm{inter}^2}{(E_{\textrm{CT}}^{\textrm{NN}}-E_{\textrm{T}})}.
\end{equation}

\subsubsection{Interchain: Slip-stacked}

\begin{equation}\label{Eq:A22}
  \langle \textrm{CT}^{+}_+; i,i+1 |\hat{H}| \textrm{FE}^+_+   \rangle = -\frac{(\alpha + \beta)}{\sqrt{2}} t_{\textrm{inter}},
\end{equation}

\begin{equation}\label{Eq:A23}
  \langle \textrm{CT}^{\pm}_+;i,j |\hat{H}| \textrm{EX}^{\pm}_+ ; i,j\rangle = -(\alpha + \beta)^2 t_{\textrm{inter}},
\end{equation}

\begin{equation}\label{Eq:A24}
  \langle \textrm{CT}^{\pm}_+;i+1,j |\hat{H}| \textrm{EX}^{\pm}_+ ; i,j\rangle = - \frac{(\alpha + \beta)^2}{2} t_{\textrm{inter}},
\end{equation}

\begin{equation}\label{Eq:A25}
  \langle \textrm{CT}^{\pm}_+;i,j-1 |\hat{H}| \textrm{EX}^{\pm}_+ ; i,j\rangle = - \frac{(\alpha + \beta)^2}{2} t_{\textrm{inter}},
\end{equation}

\begin{equation}\label{Eq:A26}
  \langle \textrm{CT}^-; i,\bar{i} |\hat{H}| \textrm{TT}^-;i,\bar{i} \rangle = \sqrt{\frac{3}{8}} t_{\textrm{inter}},
\end{equation}
and
\begin{equation}\label{Eq:A261}
  \langle \textrm{iTT}_+^-;i,j |\hat{H}| \textrm{TT}_+^- ;i,j  \rangle =  - \frac{t_\textrm{inter}^2}{(E_{\textrm{CT}}^{\textrm{NN}}-E_{\textrm{T}})}.
\end{equation}

\begin{figure}[htbp]
\centering
\includegraphics[width=0.8\linewidth]{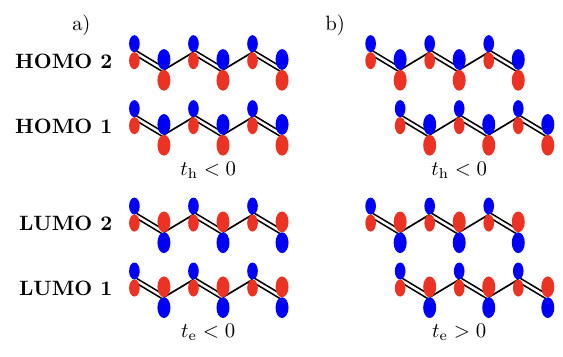}
\caption{The signs of the interchain electron- and hole-transfer integrals ($t_\textrm{e}$ and $t_\textrm{h}$, respectively) for a) vertical- and b) slip-stacked dimers. The color coding of the atomic-orbital represents their relative signs in the HOMOs and LUMOs.
The difference in the relative signs of the electron- and hole-transfer integrals accounts for the different selection rules governing interchain coupling.
Note that these signs do not include phases resulting from fermionic anticommutation.}
\label{hole_transfer}
\end{figure}

These selection rules  imply a qualitative distinction between the two conformations. 
In a vertically stacked dimer, coupling between interchain CT excitons and both the bright-dimer state and interchain triplet-pairs vanishes. However, both couplings are allowed for slip-stacked dimers. This difference in couplings arises from the  opposite signs of the hole-transfer integrals between chains, as illustrated in Fig.\ \ref{hole_transfer}.

We also observe from  Eqns.~\ref{Eq:A211} and \ref{Eq:A261}  that iTT - TT coupling is a second-order process involving a high-energy CT exciton.

\subsection{Basis state energies}\label{Se:S_energies}

The excitation energy of the  triplet-state basis state (defined in Eqn.~\ref{Eq:16}) is
$E_\textrm{T}=(V_1-E_{\textrm{GS}})$, where $V_l$ is the Ohno potential between carbon atoms with a separation of \textit{l} bonds,
\begin{equation}\label{eq:Ohno}
    V_{\textit{l}} = \frac{U}{\sqrt{1 + \left( U \epsilon_r r_{\textit{l}}/14.397 \right)^2}}.
\end{equation}

The ground state energy of an ethylene dimer is
\begin{equation}\label{Eq:A16}
    E_{\textrm{GS}} = \frac{U+V_1}{2} - \epsilon, 
\end{equation}
where
\begin{equation}
    \epsilon = \frac{1}{2}\sqrt{(U-V_1)^2+16t_d^2},
\end{equation}
\begin{equation}
        \alpha=\frac{1}{2}\sqrt{\frac{2\epsilon + U-V_1}{\epsilon}} 
\end{equation}
and $\beta=\sqrt{1-\alpha^2}$.
For realistic PPP parameters, $\alpha^2\approx\frac{2}{3}$, $\beta^2\approx\frac{1}{3}$.

The excitation energy of the Frenkel exciton basis state, $|\textrm{FE}\rangle$,  (defined in Eqn.~\ref{Eq:17}) is $E_{\textrm{FE}}=(U-E_{\textrm{GS}})$. The excitation energy of an intrachain charge-transfer exciton basis state, $|\textrm{EX}\rangle$, (defined in Eqn.~\ref{Eq:18}) is
\begin{equation}
    E_{EX}(j-i)=U+2V_1-2t_d-2E_{GS}-\tilde{V}_{|j-i|},
\label{eq:EXenergy}
\end{equation}
where the electron-hole Coulomb attraction is
\begin{equation}
    \tilde{V}_l=(V_{2l-1}+2V_{2l}+V_{2l+1})/4.
\end{equation}
The excitation energy of both the intra- and interchain triplet-pair basis states (defined in Eqns.~\ref{Eq:19} and \ref{Eq:21}, respectively) is $E_{\textrm{T}\textrm{T}}=2E_\textrm{T}$.
Finally, the excitation energy, $E_{\textrm{CT}}(i_x,j_{\bar{x}})$, of an interchain charge-transfer exciton basis state, $|\textrm{CT}\rangle$, (defined in Eqn.~\ref{Eq:20}) is similar to $E_{EX}$, except that $\tilde{V}$ extends across chains as,
\begin{equation}
    \tilde{V}_{l}=(V_{\|D,2l-1\|}+2V_{\|D,2l\|}+V_{\|D,2l+1\|})/4.
\end{equation}

\bibliography{AI_refs}